\documentclass[a4paper]{article}
\usepackage{amsmath,amssymb,amsthm,amsfonts}
\usepackage[colorlinks=true,linkcolor=black,citecolor=black,urlcolor=black]{hyperref}
\usepackage{bbm,dsfont,mathrsfs}
\usepackage{graphicx,epsfig,color,subcaption}
\usepackage[margin=3cm]{geometry}

\renewcommand{\thefootnote}{\arabic{footnote}}
\numberwithin{equation}{section}

\newcommand{\id}{{\boldsymbol{\mathbbm{1}}}}
\DeclareMathOperator{\tr}{tr}
\DeclareMathOperator{\sym}{sym}
\DeclareMathOperator{\dev}{dev}
\DeclareMathOperator{\skewp}{skew}

\DeclareMathOperator{\Curl}{Curl}
\DeclareMathOperator{\grad}{grad}

\DeclareMathOperator{\polar}{polar}

\newcommand{\B}{\mathbf}
\newcommand{\D}{\displaystyle}

\begin{document}

\title{Soliton solutions in geometrically nonlinear Cosserat micropolar elasticity with large deformations}

\author{\renewcommand{\thefootnote}{\arabic{footnote}}  
  Christian G. B\"ohmer\footnotemark[1] \ and
  Yongjo Lee\footnotemark[2] \ and
  Patrizio Neff\footnotemark[3]} 
\date{\today}

\footnotetext[1]{Christian G. B\"ohmer, Department of Mathematics, University College London, Gower Street, London, WC1E 6BT, UK, email: c.boehmer@ucl.ac.uk}

\footnotetext[2]{Yongjo Lee, Department of Mathematics, University College London, Gower Street, London, WC1E 6BT, UK, email: yongjo.lee.16@ucl.ac.uk}

\footnotetext[3]{Patrizio Neff, Fakult\"at f\"ur Mathematik, Universit\"at Duisburg-Essen, Thea-Leymann-Stra\ss e 9, 45127 Essen, Germany, email: patrizio.neff@uni-due.de} 

\date{\today}
\maketitle

\begin{abstract}
  We study the fully nonlinear dynamical Cosserat micropolar elasticity problem in three dimensions with various energy functionals dependent on the microrotation $\overline{R}$ and the deformation gradient tensor $F$. We derive a set of coupled nonlinear equations of motion from first principles by varying the complete energy functional. We obtain a double sine-Gordon equation and construct soliton solutions. We show how the solutions can determine the overall deformational behaviours and discuss the relations between wave numbers and wave velocities thereby identifying parameter values where the soliton solution does not exist. 
\end{abstract}

\mbox{}


\mbox{}


\mbox{}

\section{Introduction}

Classical elasticity is based on considering materials whose idealised material points are structureless. Any possible internal properties are neglected in the classical theory. A microcontinuum, on the other hand, is a continuous collection of deformable materials points \cite{AE1, PN2006, PN2007, PN2013, PN2014}. The characteristic aspect of the theory with microstructure is that we assume the microelements to exhibit an inner structure attached to so-called directors, which span an internal three-dimensional space. These can, for instance, rotate and deform. The most general case of this microcontinuum is the micromorphic continuum which has nine additional degrees of freedom when compared with classical elasticity theory. These additional degrees of freedom consist of 3 microrotations, 1 (micro) volume expansion and 5 (micro) shear deformations of the directors.

A special model arises when the extra degrees of freedom are reduced to rigid rotations. This means only 3 additional microrotations are considered, in addition to the classical translational deformation field. This theory is often referred to as Cosserat elasticity or micropolar elasticity and was originally proposed in full generality by the Cosserat brothers in 1909, see \cite{EC}. In some ways, their work was ahead of their time and was consequently largely forgotten for many decades. Starting from the 1950s interests in this theory increased and many advances were made since then \cite{EW1951, JE1957-1, RT1962, JE1962-1, AG1964, AE1964-1, RM1964, RT1964, HS1967, JE1967-1}.

\begin{figure}[!htb]
  \center{\includegraphics[width=0.8\textwidth]{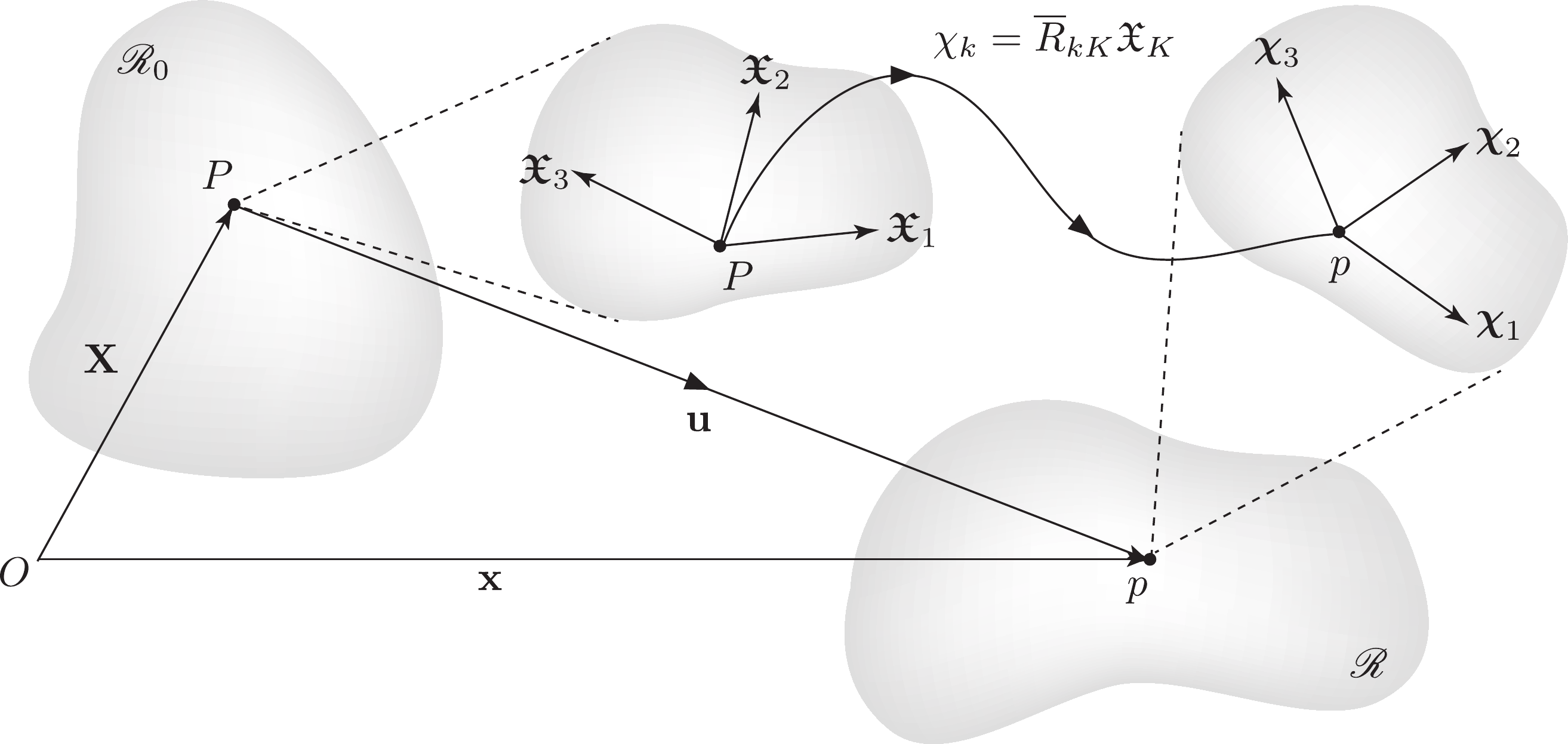}}
  \caption{The set of directors $\{\mathfrak{X},\chi\}$ determines the inner structure of the microelement with centroids positioned at $P$ and $p$ in reference configuration and spatial configuration respectively. This illustrates how the directors $\{\mathfrak{X}\}$ in the original body $\mathscr{B}_0$ undergoes microrotations under $\overline{R}$ while the original body $\mathscr{B}_0$ experiences displacements to become the deformed configuration body $\mathscr{B}$ in three-dimensional space under $\B{u}$.}
  \label{f001}
\end{figure}

We denote the microrotation by $\overline{R}=\exp(\overline{X})$, where $\overline{X}$ is a one-parameter subgroup generated by the angle $\phi=\phi(\B{x},t)$ and the deformation gradient vector is related to the displacement vector $\B{u}$ as $F=\nabla\varphi=\id+\nabla\B{u}$. The deformation gradient $F$ can be written using the polar decomposition $F=RU$ and we can express $\B{u}$ as a function of $\psi=\psi(\B{x},t)$, see Fig.~\ref{f001}.

Once one can identify and collect the relevant energy functionals for the Cosserat micropolar elasticity, the equations of motion for the system can be found through the corresponding Euler-Lagrange equation. Due to the highly nonlinear nature of the system, various attempts were made to simplify the process under relatively weak restrictions and a simple ansatz. Spinor methods were used in \cite{CB2011-1} to simplify the Euler-Lagrange equation and subsequent works appeared in \cite{CB2012-1, CB2013-1}, with an intrinsically two-dimensional model studied in \cite{CB2017-X}. An investigation in optimisation of the Cosserat shear-stretch energy in searching for the optimal Cosserat rotation is made in \cite{AF2017, AF2017-3}. In \cite{GM1986-1}, the polarity of ferromagnets gave rise to the description of the defects in order parameters as the solitary waves under the external magnetic stimuli, followed by the study in the elastic crystals as a micropolar continuum in \cite{GM1986-2}, again with the description of soliton solution for the topological defects. The reader may imagine an artificial discrete model where rotational discs connected by line and torsional springs are coupled in chiral fashion as a mechanical realisation of the model. Variants of the geometrically nonlinear Cosserat model are also used to describe lattice rotations in metal plasticity, see for instance~\cite{AF2017-2}.

In the recent paper \cite{CB2016-2}, the dynamical Cosserat model was investigated by analysing the geometrically nonlinear and coupled nature of the system, in which the linearised energy functionals are used to simplify the problem significantly. It allowed the reduction of the coupled system of PDEs to a \textit{sine-Gordon equation}, which in turn yielded a soliton-like solutions both in rotational and displacement deformations under the assumption that displacements are small while large and multiple rotations are allowed.

In this paper, we present the solutions of elastic and rotational propagation of deformations in the complete dynamical Cosserat problem. This involves the total energy functional given by
\begin{align}
  V=V_{\text{elastic}}(F,\overline{R})+V_{\text{curvature}}(\overline{R})+V_{\text{interaction}}(F,\overline{R})+V_{\text{coupling}}(F,\overline{R}).
\end{align}
We will start with exactly the same ansatz used in \cite{CB2016-2} such that the displacement deformation wave is a plane wave in the form of $\psi=g(z-vt)$ for some arbitrary function $g$ with wave speed $v$. We expect to obtain a similar system of equations but with additional nonlinear terms. It turns out that these will yield a \textit{double sine-Gordon type equation}.

The primary mathematical interests in finding the equations of motion using the variational calculus come from the fact that many terms in the energy functional contain quantities such as $\overline{R}^T\Curl\overline{R}$, $\overline{R}^T\text{polar}(F)$, or $\overline{R}^TF$ throughout the calculations. Since in general the elements $\overline{R}\in \mathrm{SO}(3)$ do not commute, their variations require a careful treatment in the calculations.

The plan of the paper is the following. In Section 2, after stating each energy functional in terms of $F$ and $\overline{R}$, we vary the total energy functional including kinetic energy. We collect terms from the variational field expressions with respect to $F$ and $\overline{R}$ in Section 3 to obtain the complete coupled system of equations of motion. It turns out that if we impose the previous restriction, i.e.~small displacements, the newly generated nonlinear coupling terms in the complete description are indeed responsible for the contribution in the additional terms of the sine-Gordon type equation, as shown in Section 4. This observation reduces the problem to solving the so-called \textit{double sine-Gordon equation} \cite{PB} of a single function of $\phi=\phi(z,t)$. In the final Section we illustrate the effects of rotational and displacement propagations in the simple model of microcontinuum with additional features of kink-antikink form of solutions and the profiles of the wave number $k$ and wave velocity $v$ relations.

\section*{Notation}

\begin{tabular}{ll}
  $\mathbbm{1}$ & identity matrix \\
  $\varphi$ & deformation vector\\
  $\phi$ & rotation angle\\
  $\mathbf{u}$ & displacement vector \\
  $\mathbf{a}$ & rotation vector \\
  $F=\nabla\varphi=\id+\nabla\mathbf{u}$ & deformation gradient \\
  $F_{ij} = \delta_{ij} + \mathbf{u}_{i,j} = \delta_{ij} + \partial_j \mathbf{u}_{i}$ & deformation gradient in index notation\\
  $\overline{R} = \exp(\overline{X})$ & rotation matrix, microrotation \\
  $\overline{X}$ & skew-symmetric matrix generating $\overline{R}$ \\
  $\epsilon_{ijk}$ & Levi-Civita symbol, $\epsilon_{123} =1 = -\epsilon_{213}$\\
  $\overline{U} = \overline{R}^T F$ & non-symmetric stretch tensor, first Cosserat deformation tensor  \\
  $F = R\,U = \polar(F) U$ & classical polar decomposition \\
  $(\Curl M)_{ij} = \varepsilon_{jrs} \partial_r M_{is}$ & matrix Curl \\
  $\sym M = (M+M^T)/2$ & symmetric part of matrix $M$\\
  $\skewp M  = (M-M^T)/2$ & skew-symmetric part of $M$\\
  $\dev M = M-\tr(M)\id/3$ & deviatoric or trace-free part of $M$ \\
  $A:B=\langle A,B\rangle=\tr(AB^T)=\tr(A^TB)$ & Frobenius product of matrices $A$ and $B$\\
  $\|X\|^2=\langle X,X\rangle=\tr(XX^T)$ & Frobenius norm of $X$
\end{tabular}

\section{The complete dynamical Cosserat problem}

We introduce each energy functional for the full treatment of the geometrically nonlinear Cosserat problem in three-dimensional space. We will subtract the kinetic energies from relevant energy functionals before deriving the equations of motion. First, the energy functional for elastic deformations is
\begin{equation}\label{2.1}
  V_{\text{elastic}}(F,\overline{R})=\mu\left\|\text{sym}\ \overline{R}^{T}F-\id\right\|^{2}+\frac{\lambda}{2}\bigl[\tr(\text{sym} (\overline{R}^{T}F)-\id)\bigr]^{2}
\end{equation}
where $\lambda$ and $\mu$ are the standard Lam\'{e} parameters. The microrotations are governed by the energy functional $V_{\text{curvature}}$ defined by
\begin{equation}\label{2.2}
  \begin{split}
    V_{\text{curvature}}(\overline{R})&=\kappa_{1}\left\|\text{dev}\;\text{sym}(\overline{R}^{T}\Curl\overline{R})\right\|^{2}+\kappa_{2}\left\|\text{skew}(\overline{R}^{T}\Curl\overline{R})\right\|^{2}\\
    &\qquad+\kappa_{3}\Bigl[\tr(\overline{R}^{T}\Curl\overline{R})\Bigr]^{2}
  \end{split}
\end{equation}
where $\kappa_i$ are the elastic constants for the microrotations. 

An interaction between elastic displacements and microrotations is described by the irreducible parts of the elastic deformations and microrotations, such as $\overline{R}^TF-\id$ and $\overline{R}^T\Curl\overline{R}$ respectively to form the energy functional $V_{\text{interaction}}(F,\overline{R})$ defined by
\begin{equation}\label{2.3}
  \begin{split}
    V_{\text{interaction}}(F,\overline{R})&=\chi_{1}\tr(\overline{R}^{T}\Curl\overline{R})\tr(\overline{R}^{T}F)\\
    &\qquad+\chi_{3}\langle\text{dev}\;\text{sym}(\overline{R}^{T}\Curl\overline{R}),\;\text{dev}\;\text{sym}(\overline{R}^{T}F-\id)\rangle.
  \end{split}
\end{equation}
where $\chi_{1}$ and $\chi_{3}$ are the coupling constants.

Finally, we will consider the Cosserat coupling term which is given by
\begin{equation}\label{2.4}
  V_{\text{coupling}}(F,\overline{R})=\mu_{c}\left\|\overline{R}^{T}\ \text{polar}(F)-\id\right\|^{2}
\end{equation}
where $\mu_c$ is the Cosserat couple modulus.

The variations of the complete energy functional are quite involved. All required results are stated explicitly in Appendix~\ref{sec:app}. Gathering all the variational terms (\ref{a2.10}), (\ref{a2.18}), (\ref{a2.20}) and (\ref{a2.24}), we will obtain the complete variational functional of the theory for the dynamical case
\begin{equation}
  \label{2.25}
  \delta V(F,\overline{R})=\delta V_{\text{coupling}}(F,\overline{R})+\delta V_{\text{interaction}}(F,\overline{R})+\delta V_{\text{elastic}}(F,\overline{R})+\delta V_{\text{curvature}}(\overline{R})
\end{equation}
where
\begin{align}
  \label{2.26}
  \nonumber
  \delta V_{\rm elastic}(F,\overline{R})&=\Big[\mu(\overline{R}F^{T}\overline{R}+F)-(2\mu+3\lambda)\overline{R}+\lambda \tr(\overline{R}^{T}F)\overline{R}\Big]:\delta F\\ \nonumber
  &\qquad+\Big[\mu F\overline{R}^{T}F-(2\mu+3\lambda)F+\lambda\tr(\overline{R}^{T}F)F\Big]:\delta \overline{R}+\rho\ddot{u}\;\delta u\\ \nonumber
  \delta V_{\text{curvature}}(\overline{R})&=\Big[(\kappa_{1}-\kappa_{2})\Big((\Curl\overline{R})\overline{R}^{T}(\Curl(\overline{R}))+\Curl\Big[\overline{R}(\Curl\overline{R})^{T}\overline{R}\Big]\Big)+(\kappa_{1}+\kappa_{2})\Curl\Big[\Curl\overline{R}\Big]\\ \nonumber
    &-\left(\frac{\kappa_{1}}{3}-\kappa_{3}\right)\Big(4\tr(\overline{R}^T\Curl\overline{R})\Curl(\overline{R}) -2\overline{R}\Big(\grad\Big(\tr[\overline{R}^{T}\Curl\overline{R}]\Big)\Big)^{\star}+2\rho_{\text{rot}}\ddot{\overline{R}}\Big]:\delta\overline{R}\\ \nonumber
  \delta V_{\text{interaction}}(F,\overline{R})&=\left\{\Big(\chi_{1}-\frac{\chi_{3}}{3}\Big)\Big(2\tr(\overline{R}^{T}F)\Curl(\overline{R})+\tr(\overline{R}^{T}\Curl\overline{R})F-\overline{R}\Big[\grad\Big(\tr[\overline{R}^{T}F]\Big)\Big]^{\star}\Big)\right.\\ \nonumber
  &+\left.\frac{\chi_{3}}{2}\Big(\Curl(F)+(\Curl(\overline{R}))\overline{R}^{T}F+F\overline{R}^{T}(\Curl(\overline{R}))+\Curl(\overline{R}F^{T}\overline{R})\Big)\right\}:\delta \overline{R}\\ \nonumber
  &+\left\{\chi_{1}\tr(\overline{R}^{T}\Curl\overline{R})\overline{R}+\frac{\chi_{3}}{2}\Big(\Curl(\overline{R})+\overline{R}(\Curl(\overline{R}))^{T}\overline{R}\Big)-\frac{\chi_{3}}{3}\tr(\overline{R}^{T}\Curl(\overline{R}))\overline{R}\right\}:\delta F\\
  \delta V_{\text{coupling}}(F,\overline{R})&=-2\mu_{c}\overline{R}:\delta\overline{R}-\frac{2\mu_{c}}{\det(Y)}\left[RY(R^{T}\overline{R}-\overline{R}^{T}R)Y\right]:\delta F.
\end{align}                                                                                                                                 
The next step will be collecting the various expressions with respect to $F$ and $\overline{R}$ to construct the field equations.

\section{Equations of motion and solutions}

\subsection{Displacements and rotations in one axis}

Let us assume that the points in our continuum can only experience rotations about one axis, say the $z$-axis, which means we can choose
\begin{equation}\label{3.1}
  \overline{R}=
  \begin{pmatrix}
    \cos\phi&-\sin\phi &0\\
    \sin\phi&\cos\phi &0\\
    0&0&1
  \end{pmatrix}.
\end{equation}
The variation of this is simply
\begin{equation}
  \delta\overline{R}=
  \begin{pmatrix}
    -\sin\phi\;\delta\phi&-\cos\phi\;\delta\phi&0\\
    \cos\phi\;\delta\phi&-\sin\phi\;\delta\phi&0\\
    0&0&0
  \end{pmatrix}.
\end{equation}

In principle, the rotational and elastic waves can be either longitudinal or transverse in each case, hence four different combinations are possible. Here, we consider solutions in which both waves are longitudinal about the same axis, the $z$-axis in this case, so that we can write $\psi=\psi(t,z)$ and $\phi=\phi(t,z)$.
\begin{equation}\label{3.2}
  \B{u}=
  \begin{pmatrix}
    0\\
    0\\
    \psi(z,t)
  \end{pmatrix}
  ,\quad
  \nabla \B{u}=
  \begin{pmatrix}
    0&0&0\\
    0&0&0\\
    0&0&\partial_{z}\psi(z,t)
  \end{pmatrix} \,.
\end{equation}

Further, we collect the relevant terms with respect to $F$ and $\overline{R}$ separately. Unlike the case of $\delta\overline{R}$, in which the variational kinetic term is readily written with respect to $\overline{R}$, the variational kinetic term from the interaction energy functional is written with respect to $\delta \B{u}$. But the variation with respect to $F$ can be restated as the variation with respect to $\B{u}$, hence with respect to $\psi$ as we will see shortly.

Collecting terms for $\delta F$ from (\ref{2.26}) gives
\begin{multline}
  \begin{pmatrix}
    A_{11}&A_{12}&0\\
    -A_{12}&A_{11}&0\\
    0&0&A_{33}
  \end{pmatrix} :=
  \left[\mu\left(\overline{R}F^T\overline{R}+F\right)-(2\mu+3\lambda)\overline{R}+\lambda\tr(\overline{R}^TF)\overline{R}\right] \\[1ex]
  +\left[\chi_1\tr\left(\overline{R}^T\Curl\overline{R}\right)\overline{R}+\frac{\chi_3}{2}\left(\Curl\overline{R}+\overline{R}(\Curl\overline{R})^T\overline{R}\right)-\frac{\chi_3}{3}\tr\left(\overline{R}^T\Curl\overline{R}\right)\overline{R}\right]\\[1ex]
  +\frac{2\mu_c}{\text{det}\;Y}\left[RY\left(\overline{R}^TR-R^T\overline{R}\right)Y\right]\,,
  \label{3.4}
\end{multline}
where
\begin{equation}\label{3.5}
  \begin{split}
    A_{11}&=\frac{1}{3}\cos\phi\Bigl[6(\lambda+\mu)(-1+\cos\phi)+(6\chi_1+\chi_3)\partial_z\phi+3\lambda\partial_z\psi\Bigr],\\
    A_{12}&=-\frac{1}{3}\sin\phi\Bigl[-6\lambda-6\mu+6\lambda\cos\phi+6\mu\cos\phi-6\mu_c+(6\chi_1+\chi_3)\partial_z\phi+3\lambda\partial_z\psi\Bigr],\\
    A_{33}&=2\lambda(-1+\cos\phi)+\left(2\chi_1-\frac{2\chi_3}{3}\right)\partial_z\phi+(\lambda+2\mu)\partial_z\psi.
  \end{split}
\end{equation}
Now, the terms which appear in the variation with respect to $F$ can be transformed into the variation with respect to $\nabla \B{u}$, for any matrix $\B{A}$, as follow.
\begin{equation}
  \label{3.6}
  \B{A}:\delta F=A_{ij}\delta F_{ij}
  \longrightarrow -\partial_jA_{ij}\delta u_i = 
  -(\partial_1A_{31}+\partial_2A_{32}+\partial_3A_{33})\delta\psi,
\end{equation}
up to a boundary term. In this case, the contribution comes only from $A_{33}$ and we obtain
\begin{equation}\label{3.7}
  \left[2\lambda\sin\phi\;\partial_z\phi-\left(2\chi_1-\frac{2\chi_3}{3}\right)\partial_{zz}\phi-(\lambda+2\mu)\partial_{zz}\psi\right]\delta\psi.
\end{equation}
We now include the kinetic variational term $\rho\ddot{u}\;\delta u=\rho\partial_{tt}\psi\;\delta\psi$ to obtain the equation of motion for $F$
\begin{equation}\label{3.8}
  -\lambda \left( \partial_{zz}\psi-2 \partial_{z}\phi \sin \phi \right)-2 \mu  \partial_{zz}\psi +\rho  \partial_{tt}\psi+\frac{2}{3} (\chi_{3}-3\chi_{1}) \partial_{zz}\phi=0\;.
\end{equation}

In the same way, we collect terms for $\delta\overline{R}$ to obtain
\begin{multline}
  \begin{pmatrix}
    B_{11}&B_{12}&0\\
    -B_{12}&B_{11}&0\\
    0&0&B_{33}
  \end{pmatrix} :=
  2\rho_{\text{rot}}\ddot{\overline{R}}
  +\mu F\overline{R}^{T}F-(2\mu+3\lambda)F+\lambda \ \textrm{tr}(\overline{R}^{T}F)F
  -2\mu_{c}R \\[1ex]
  +(\kappa_{1}-\kappa_{2})\Big[(\textrm{Curl}\overline{R})\overline{R}^{T}(\textrm{Curl}\overline{R})+\textrm{Curl}\left(\overline{R}(\textrm{Curl}\overline{R})^{T}\overline{R}\right)\Bigr]+(\kappa_{1}+\kappa_{2})\Bigl[\textrm{Curl}(\textrm{Curl}\overline{R})\Bigr]\\
  -\left(\frac{\kappa_{1}}{3}-\kappa_{3}\right)\Big[4\textrm{tr}(\overline{R}^{T}\ \textrm{Curl}\overline{R})\textrm{Curl}\overline{R}-2\overline{R}\Big(\textrm{grad}\Big[\textrm{tr}(\overline{R}^{T}\ \textrm{Curl}\overline{R})\Big]\Big)^{\star}\Big]
  \\[1ex]
+\left(\chi_{1}-\frac{\chi_{3}}{3}\right)\left(2\tr(\overline{R}^{T}F)\Curl\overline{R}+\textrm{tr}(\overline{R}^{T} \ \textrm{Curl}\overline{R})F-\overline{R}\Big[\textrm{grad}(\textrm{tr}(F\overline{R}^{T}))\Big]^{\star}\right)\\
  +\frac{\chi_{3}}{2}\left(\textrm{Curl}F+(\textrm{Curl}\overline{R})\overline{R}^{T}F+F\overline{R}^{T}(\textrm{Curl}\overline{R})+\textrm{Curl}( \overline{R}F^{T}\overline{R})\right)
\label{3.9}
\end{multline}
where
\begin{align}\label{3.10}\nonumber
  B_{11}=&-2(\lambda+\mu+\mu_c)+6\chi_1\cos^2\phi\;\partial_z\phi+\lambda\partial_z\psi\\ \nonumber
  &+\frac{1}{3}\cos\phi\Bigl[3(2\lambda+\mu)-6\rho_{\text{rot}}(\partial_t\phi)^2+(\kappa_1-3\kappa_2+24\kappa_3)(\partial_z\phi)^2+2(3\chi_1-\chi_3)\partial_z\phi(1+\partial_z\psi)\Bigr]\\  \nonumber
  &+\frac{1}{3}\sin\phi\Bigl[-6\rho_{\text{rot}}\partial_{tt}\phi+2(\kappa_1+6\kappa_3)\partial_{zz}\phi+(3\chi_1-\chi_3)\partial_{zz}\psi\Bigr],
  \\ \nonumber
  B_{12}=&\frac{1}{3}\sin\phi\Bigl[3\mu+6\rho_{\text{rot}}(\partial_t\phi)^2-(\kappa_1-3\kappa_2+24\kappa_3)(\partial_z\phi)^2-2(3\chi_1-\chi_3)\partial_z\phi(1+\partial_z\psi)\Bigr]\\ \nonumber
  &+\cos\phi\Bigl[-6\chi_1\sin\phi\;\partial_z\phi+\frac{1}{3}\left(-6\rho_{\text{rot}}\partial_{tt}\phi+2(\kappa_1+6\kappa_3)\partial_{zz}\phi+(3\chi_1-\chi_3)\partial_{zz}\psi\right)\Bigr],\\ \nonumber
  B_{33}=&-2\mu_c+2\lambda\cos\phi(1+\partial_z\psi)+\frac{1}{3}(1+\partial_z\psi) \Bigl[(6\chi_1-2\chi_3)\partial_z\phi+3(-2\lambda-\mu+(\lambda+\mu)\partial_z\psi)\Bigr]\,.
\end{align}
Applying $B:\delta\overline{R}$ gives 
\begin{equation}\label{3.11}
  B:\delta\overline{R}=\tr\left[B^T\delta\overline{R}\right]=-(2B_{11}\sin\phi+2B_{12}\cos\phi)\delta\phi
\end{equation}
which is
\begin{equation}\label{3.12}
  \begin{split}
    &\Bigl[4(\lambda+\mu+\mu_c)\sin\phi-2(\lambda+\mu)\sin 2\phi-2\lambda\sin\phi\;\partial_z\psi+4\rho_{\text{rot}}\partial_{tt}\phi\\
      &\qquad\qquad\qquad\qquad\qquad-4\left(\frac{\kappa_1}{3}+2\kappa_3\right)\partial_{zz}\phi-2\left(\chi_1-\frac{\chi_3}{3}\right)\partial_{zz}\psi\Bigr]\delta\phi.
  \end{split}
\end{equation}
Therefore, from (\ref{3.8}) and (\ref{3.12}), we obtain two equations of motion by varying the total energy functional with respect to $F$ and $\overline{R}$, respectively, as follows
\begin{subequations}\label{3.13}
  \begin{align}
    &-(\lambda+\mu+\mu_{c}) \sin\phi+\frac{1}{2} (\lambda +\mu )\sin 2 \phi+\frac{1}{2}\lambda\sin\phi \partial_{z}\psi-\rho_{\rm rot} \partial_{tt}\phi
    \nonumber \\
    &\;\;\;\quad\qquad\qquad\qquad\qquad+\left(\frac{\kappa_{1}}{3}+2 \kappa_{3}\right) \partial_{zz}\phi+\left(\frac{\chi_{1}}{2}-\frac{\chi_{3}}{6}\right)\partial_{zz} \psi=0\\ 
    &-\lambda \left( \partial_{zz}\psi-2 \partial_{z}\phi \sin \phi \right)-2 \mu  \partial_{zz}\psi +\rho  \partial_{tt}\psi+\frac{2}{3} (\chi_{3}-3\chi_{1}) \partial_{zz}\phi=0.
  \end{align}
\end{subequations}
These can be written in component form as
\begin{equation}\label{3.13-1}
  \begin{pmatrix}
    \partial_{tt}\phi\\
    \partial_{tt}\psi
  \end{pmatrix}
  =
  \B{M}
  \begin{pmatrix}
    \partial_{zz}\phi\\
    \partial_{zz}\psi
  \end{pmatrix}
  +
  \begin{pmatrix}
    0&\frac{\lambda\sin\phi}{2\rho_{\text{rot}}}\\
    -\frac{2\lambda\sin\phi}{\rho}&0
  \end{pmatrix}
  \begin{pmatrix}
    \partial_z\phi\\
    \partial_z\psi
  \end{pmatrix}
  -\frac{(\lambda+\mu+\mu_c)}{\rho_{\text{rot}}}
  \begin{pmatrix}
    \sin\phi\\0
  \end{pmatrix}
  +
  \frac{\lambda+\mu}{2\rho_{\text{rot}}}
  \begin{pmatrix}
    \sin2\phi\\0
  \end{pmatrix}
\end{equation}
where
\begin{equation}\label{3.13-2}
  \B{M}=
  \begin{pmatrix}
    (\kappa_1+6\kappa_3)/3\rho_{\text{rot}}&(3\chi_1-\chi_3)/6\rho_{\text{rot}}\\
    2(3\chi_1-\chi_3)/3\rho&(\lambda+2\mu)/\rho
  \end{pmatrix}.
\end{equation}
From this, we can see immediately that we will recover the result obtained in \cite{CB2016-2} if we assume the linearised energy functionals which lead to the approximations such as $\lambda\phi\ll1$ and $\mu\phi\ll1$, while the matrix elements $\B{M}$ remain unchanged.

The revised results of \cite{GM1986-1} were stated in \cite{AE1}, in which case the longitudinal wave is expressed as $U(x,t)$ along the $x$ axis with the rotational deformation $\phi(x,t)$ about $x$ axis. The equations of motion are described as a system of coupled expressions,
\begin{equation}\label{3.13-3}
  \begin{pmatrix}
    \partial_{tt}\phi\\
    \partial_{tt}U
  \end{pmatrix}
  =
  \B{N}
  \begin{pmatrix}
    \partial_{xx}\phi\\
    \partial_{xx}U
  \end{pmatrix}
  +
  \begin{pmatrix}
    0&\frac{2\lambda'\sin\phi}{\rho_0J}\\
    -\frac{2(\lambda'+2\mu'+\kappa')\sin\phi}{\rho_0}&0
  \end{pmatrix}
  \begin{pmatrix}
    \partial_x\phi\\
    \partial_xU
  \end{pmatrix}
  +\frac{2\lambda'}{\rho_0J}
  \begin{pmatrix}
    \sin\phi\\0
  \end{pmatrix}
  +
  \frac{2\lambda'+\mu'}{\rho_0J}
  \begin{pmatrix}
    \sin2\phi\\0
  \end{pmatrix}
\end{equation}
where $\alpha, \lambda',\mu',\kappa'$ are isotropic material moduli used in \cite{AE1} and
\begin{equation}\label{3.13-2b}
  \B{N}=
  \begin{pmatrix}
    \frac{\alpha}{\rho_0J}&0\\
    0&\frac{\lambda'+2\mu'+\kappa'}{\rho_0}
  \end{pmatrix}.
\end{equation}
Since the matrix $\B{N}$ is diagonal, we do not have second order coupling terms in the equations of motion. And under the small displacement limit, the system is readily solvable using the conventional method for the one-dimensional d'Alembert's solution subject to the appropriate boundary conditions.

\subsection{Solution for the double sine-Gordon equation}

We assume that the elastic and rotational waves propagate with the same wave speed $v$ and $\psi=g(z-vt)$, so that $\psi$ satisfies $\partial_{tt}\psi=v^2\partial_{zz}\psi$. Without this assumption we are not able to construct a solution. Now, we define $v^2_{\text{rot}}=M_{11}$ and $v^2_{\text{elas}}=M_{22}$. Then (\ref{3.13}b) becomes
\begin{equation}\label{3.14}
  g''(z-vt)=\partial_{zz}\psi=\frac{M_{21}}{v^2-v^2_{\text{elas}}}\partial_{zz}\phi-\frac{2\lambda}{\rho(v^2-v^2_{\text{elas}})}\sin\phi\;\partial_z\phi.
\end{equation}
Integrating with respect to $z$ once gives
\begin{equation}\label{3.15}
  g'(z-vt)=\partial_z\psi=\frac{M_{21}}{v^2-v^2_{\text{elas}}}\partial_z\phi+\frac{2\lambda}{\rho(v^2-v^2_{\text{elas}})}\cos\phi
\end{equation}
in which we set the constant of integration to zero by imposing the boundary condition $\psi=\partial_z\psi=0$ as $z\to\pm\infty$. Substituting (\ref{3.14}) and (\ref{3.15}) into the remaining equation of motion (\ref{3.13}a) gives
\begin{equation}\label{3.16}
  \begin{split}
    &\partial_{tt}\phi-\left[v^2_{\text{rot}}+\frac{M_{12}M_{21}}{v^2-v^2_{\text{elas}}}\right]\partial_{zz}\phi-\frac{\lambda}{2(v^2-v^2_{\text{elas}})}\left[\frac{M_{21}}{\rho_{\text{rot}}}-\frac{4M_{12}}{\rho}\right]\sin\phi\;\partial_z\phi\\
    &\qquad\qquad\qquad\qquad\qquad+\frac{(\lambda+\mu+\mu_c)}{\rho_{\text{rot}}}\sin\phi-\left[\frac{\lambda^2}{2\rho_{\text{rot}}\rho(v^2-v^2_{\text{elas}})}+\frac{\lambda+\mu}{2\rho_{\text{rot}}}\right]\sin2\phi=0.
  \end{split}
\end{equation}
Moreover, if we rescale $z$ as
\begin{equation}\label{3.17}
  z=\left(v^2_{\text{rot}}+\frac{M_{12}M_{21}}{v^2-v^2_{\text{elas}}}\right)^{1/2}\hat{z}
\end{equation}
then (\ref{3.16}) reduces to, the so-called \textit{double sine-Gordon equation}
\begin{equation}\label{3.18}
  \partial_{tt}\phi-\partial_{\hat{z}\hat{z}}\phi+m^2\sin\phi+\frac{b}{2}\sin 2\phi=0,
\end{equation}
where
\begin{equation}\label{3.19}
  m^2=\frac{(\lambda+\mu+\mu_c)}{\rho_{\text{rot}}}\qquad\text{and}\qquad b=-\frac{1}{\rho_{\text{rot}}}\left[\frac{\lambda^2}{\rho(v^2-v^2_{\text{elas}})}+(\lambda+\mu)\right].
\end{equation}
The apparent singularity in $b$ as $v^2$ approaches $v^2_{\text{elas}}$ can be removed if we make the further transformation on $v$ as 
\begin{equation}\label{3.20}
  v\longrightarrow\left(v^2_{\text{rot}}+\frac{M_{12}M_{21}}{v^2-v^2_{\text{elas}}}\right)^{1/2}\hat{v}.
\end{equation}
We note that this transformation on $v$ would not change our assumption on $\psi$ along with the rescaling on $z$, since $\partial_{tt}\psi=v^2\partial_{zz}\psi$ implies $\partial_{tt}\psi=\hat{v}^2\partial_{\hat{z}\hat{z}}\psi$.

The general solution of (\ref{3.18}) is given in \cite{PB} as
\begin{equation}\label{3.22}
  \phi=2\;\text{arcsin}(X)
\end{equation}
where
\begin{equation}\label{3.23}
  X=\frac{u}{\sqrt{1+\frac{1}{2}u^2\left(1+\frac{b}{m^2+b}\right)+\frac{1}{16}u^4\left(1-\frac{b}{m^2+b}\right)^2}}
\end{equation}
in which $u$ must satisfy two conditions
\begin{equation}\label{3.24}
  \begin{split}
    \partial_{tt}u-\partial_{zz}u+(m^2+b)u&=0 \,,\\
    (\partial_t u)^2 + (\partial_z u)^2 + (m^2+b)u^2&=0\,.
  \end{split}
\end{equation}
The simplest solution is of the form with
\begin{equation}\label{3.25}
  u=\exp\left[\sqrt{\frac{m^2+b}{1-\hat{v}^2}}\;(\hat{z}-\hat{v}t)\right]\;.
\end{equation}

\begin{figure}[!htb]
  \center{\includegraphics[scale=0.25]{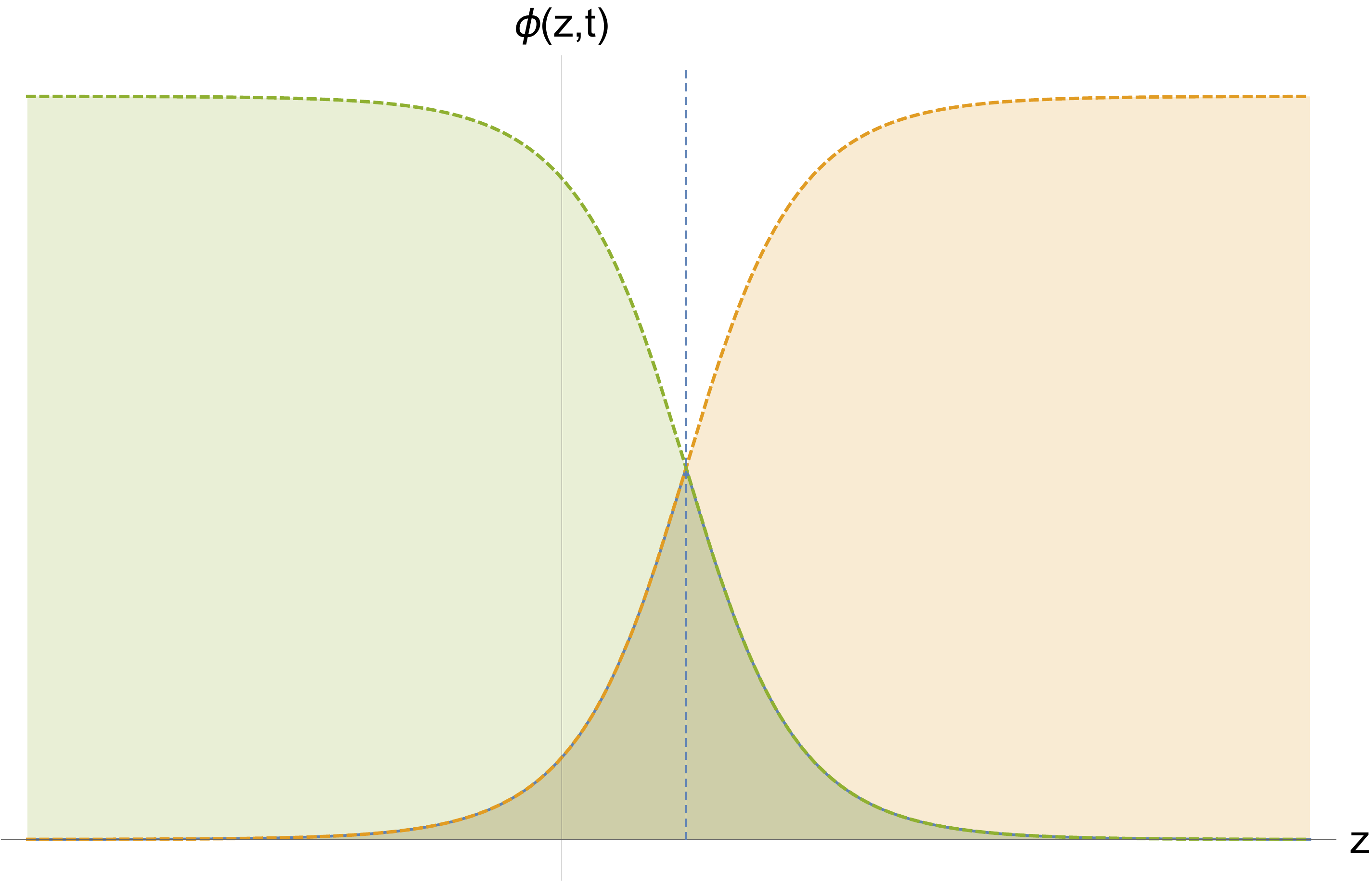}}
  \caption{Two branches of solution $\phi=4\;\text{arctan}\;e^{\pm k(z-vt)\pm\delta}$ of (\ref{3.34}) are plotted where the orange solution is for $+k$ and green is for $-k$ solution. These two branches meet at $z=\ln 4/(2k)+vt$ as indicated by the blue vertical dashed line. The overlap is essentially the solution of the form $\phi=2\;\text{arcsin}(X)$ as in (\ref{3.22}).
    We set $k=1.5$, $v=0.1$ at $t=7.0$.}
  \label{f002}
\end{figure}

Now, we can write the solution $\phi$ using the identity $\text{arcsin}(x)=2\;\text{acrtan}\left(\frac{x}{1+\sqrt{1-x^2}}\right)$ to obtain
\begin{equation}\label{3.26}
  \phi=
  \begin{cases}
    4\;\text{arctan}\Bigl[\frac{1}{2}e^{\sqrt{\frac{m^2+b}{1-\hat{v}^2}}(\hat{z}-\hat{v}t)}\Bigr]&\text{if}\quad e^{2\sqrt{\frac{m^2+b}{1-\hat{v}^2}}(\hat{z}-\hat{v}t)}<4,\\
    4\;\text{arctan}\Bigl[2e^{-\sqrt{\frac{m^2+b}{1-\hat{v}^2}}(\hat{z}-\hat{v}t)}\Bigr]
    &\text{if}\quad e^{2\sqrt{\frac{m^2+b}{1-\hat{v}^2}}(\hat{z}-\hat{v}t)}>4.
  \end{cases}
\end{equation}
This solution corresponds to the \textit{kink} and \textit{antikink} solutions of $\phi$ and the bifurcation into these two branches from the original solution (\ref{3.22}) arises quite naturally in translating the solution in terms of arcsin into arctan functions, see Fig.~\ref{f002}.

Next, we would like to put the rescaled variables $\{\hat{z},\hat{v}\}$ back to the original variables $\{z,v\}$. In~\cite{CB2016-2}, we obtained
\begin{equation}\label{3.27}
  \phi_0=4\;\text{arctan}\;e^{\pm k_0(z-vt)\pm\delta}
\end{equation}
where $\phi_0$ is the rotational propagation solution based on the linearised energy functionals  with corresponding $k_0$ and $m_0$ given by
\begin{equation}\label{3.28}
  k_0^2=\frac{v^2_{\text{elas}}-v^2}{v^4-\tr(\B{M})v^2+\text{det}(\B{M})}m_0^2,\qquad\qquad m_0^2=\frac{\mu_c}{\rho_{\text{rot}}}.
\end{equation}
Now, consider the quantity
\begin{equation}\label{3.29}
  \pm\sqrt{\frac{m^2_0}{1-\hat{v}^2}}\;(\hat{z}-\hat{v}t)\pm\delta
\end{equation}
for $\delta=\ln\frac{1}{2}$. We would like to see if this agrees with the argument of the exponential in (\ref{3.27}). This can be done if we apply the reverse rescaling (\ref{3.17}) of $z$ and inverse transformation (\ref{3.20}) of $v$. After some calculations, we obtain
\begin{equation}\label{3.30}
  \begin{split}
    \pm\sqrt{\frac{m^2_0}{1-\hat{v}^2}}\;(\hat{z}-\hat{v}t)\pm\delta&=\pm\sqrt{\frac{m^2_0}{1-\frac{v^2}{v^2_{\text{rot}}+\frac{M_{12}M_{21}}{v^2-v^2_{\text{elas}}}}}}\;\frac{1}{\sqrt{v^2_{\text{rot}}+\frac{M_{12}M_{21}}{v^2-v^2_{\text{elas}}}}}(z-vt)\pm\delta\\
    &=\pm k_0(z-vt)\pm\delta\;.
  \end{split}
\end{equation}
Hence, we can express the solution of $\phi_0$ in terms of rescaled variables $\{\hat{z},\hat{v}\}$ or the original variables $\{z,v\}$ with $k_0$ of (\ref{3.28}) and find
\begin{equation}\label{3.31}
  \phi_0=4\;\text{arctan}\;e^{\pm k_0(z-vt)\pm\delta}=4\;\text{arctan}\;e^{\pm\sqrt{\frac{m^2_0}{1-\hat{v}^2}}\;(\hat{z}-\hat{v}t)\pm\delta}.
\end{equation}

For the current case, by following the same reasoning we find that the rescaled variables and original variables are interchangeable by the expression
\begin{equation}\label{3.32}
  \pm k(z-vt)\pm\delta=\pm\sqrt{\frac{m^2+b}{1-\hat{v}^2}}\;(\hat{z}-\hat{v}t)\pm\delta
\end{equation}
where
\begin{equation}\label{3.33}
  k^2=\frac{v^2_{\text{elas}}-v^2}{v^4-\tr(\B{M})v^2+\text{det}(\B{M})}(m^2+b),\qquad\qquad m^2=\frac{\lambda+\mu+\mu_c}{\rho_{\text{rot}}}.
\end{equation}
Therefore, we can write the solution (\ref{3.26}) of $\phi$ in terms of $z$ and $v$ as
\begin{equation}\label{3.34}
  \phi=4\;\text{arctan}\;e^{\pm k(z-vt)\pm\delta}
\end{equation}
with $\delta=\ln\frac{1}{2}$.

We must notice that the matrix $\B{M}$ used in (\ref{3.33}) and (\ref{3.28}) is the same (\ref{3.13-2}). The Lam\'{e} parameters $\lambda$ and $\mu$ are brought into play in the fully nonlinear case through the quantity $m^2$, while those parameters are missing in $m^2_0$ when considering the approximations $\lambda\phi\ll 1$ and $\mu\phi\ll 1$. Consequently, we have to treat a more complicated form of $k$ with an additional contribution from $b$. And it is clear that we can recover the solution (\ref{3.31}) if we apply the restrictions $\lambda\phi\ll 1$ and $\mu\phi\ll 1$, which will effectively lead to $b=0$ and $m\to m_0$.

For $\psi$, first we write $X$ (hence $u$) in terms of $z$ and $v$.
\begin{equation}\label{3.35}
  X=\frac{u}{\sqrt{1+\frac{1}{2}u^2\left(1+\frac{b}{m^2+b}\right)+\frac{1}{16}u^4\left(1-\frac{b}{m^2+b}\right)^2}}\;,\qquad\quad u=e^{\pm k_(z-vt)\pm\delta}\;.
\end{equation}
Plugging (\ref{3.34}) into (\ref{3.14}) gives,
\begin{align}\label{3.36}
  g''(z-vt)&=\frac{4M_{21}k^2}{v^2_{\text{elas}}-v^2}\frac{e^{\pm k(z-vt)\pm\delta}(e^{2(\pm k(z-vt)\pm\delta)}-1)}{(e^{2(\pm k(z-vt)\pm\delta)}+1)^2}\\ \nonumber
  &+\frac{2\lambda}{\rho(v^2-v^2_{\text{elas}})}\frac{\pm k(m^2+b)^2\left(64e^{6(\pm k(z-vt)\pm\delta)}m^4-1024e^{2(\pm k(z-vt)\pm\delta)}(m^2+b)^2\right)}{\left(e^{4(\pm k(z-vt)\pm\delta)}m^4+16(m^2+b)^2+8e^{2(\pm k(z-vt)\pm\delta)}(m^2+b)(m^2+2b)\right)^2}.
\end{align}
If we put $s=z-vt$, then this becomes a second-order ordinary differential equation for $g(s)$. We integrate twice with respect to $s$ using the boundary conditions $\psi'(\pm\infty,t)=\psi(\pm\infty,t)=0$ to obtain
\begin{equation}\label{3.37}
  \psi=\frac{4M_{21}}{v^2-v^2_{\text{elas}}}\text{arctan}\;e^{\pm k(z-vt)\pm\delta}+\frac{4\lambda}{\rho k(v^2-v^2_{\text{elas}})}\sqrt{1+\frac{m^2}{b}}\;\text{acrtanh}(Y)+C
\end{equation}
where
\begin{equation}\label{3.38}
  Y=
  \begin{cases}
    \D{\frac{8b^2+12bm^2+m^4(\frac{1}{4}e^{2k(z-vt)}+4)}{8\sqrt{b}\;(m^2+b)^{3/2}}}&\text{if}\quad e^{2k(z-vt)}<4,\\
    \D{\frac{8b^2+12bm^2+m^4(4e^{-2k(z-vt)}+4)}{8\sqrt{b}\;(m^2+b)^{3/2}}}&\text{if}\quad e^{2k(z-vt)}>4.
  \end{cases}
\end{equation}
The constant $C$ is
\begin{equation}\label{3.38-1}
  C=-\frac{4\lambda}{\rho k(v^2-v^2_{\text{elas}})}\sqrt{1+\frac{m^2}{b}}\;\text{acrtanh}\left(\frac{8b^2+12bm^2+4m^4}{8\sqrt{b}\;(m^2+b)^{3/2}}\right).
\end{equation}
Using the restriction $\lambda\phi\ll 1$ and $\mu\phi\ll 1$, these solutions reduce to the one we obtained in \cite{CB2016-2}
\begin{equation}\label{3.39}
  \begin{split}
    \phi_0&=4\;\text{arctan}\;e^{\pm k_0(z-vt)\pm\delta}\\
    \psi_0&=\frac{4M_{21}}{v^2-v^2_{\text{elas}}}\text{arctan}\;e^{\pm k_0(z-vt)\pm\delta}\;.
  \end{split}
\end{equation}

\begin{figure}[!htb]
  \begin{tabular}{cc}
    \parbox{3in}{\includegraphics[scale=0.6]{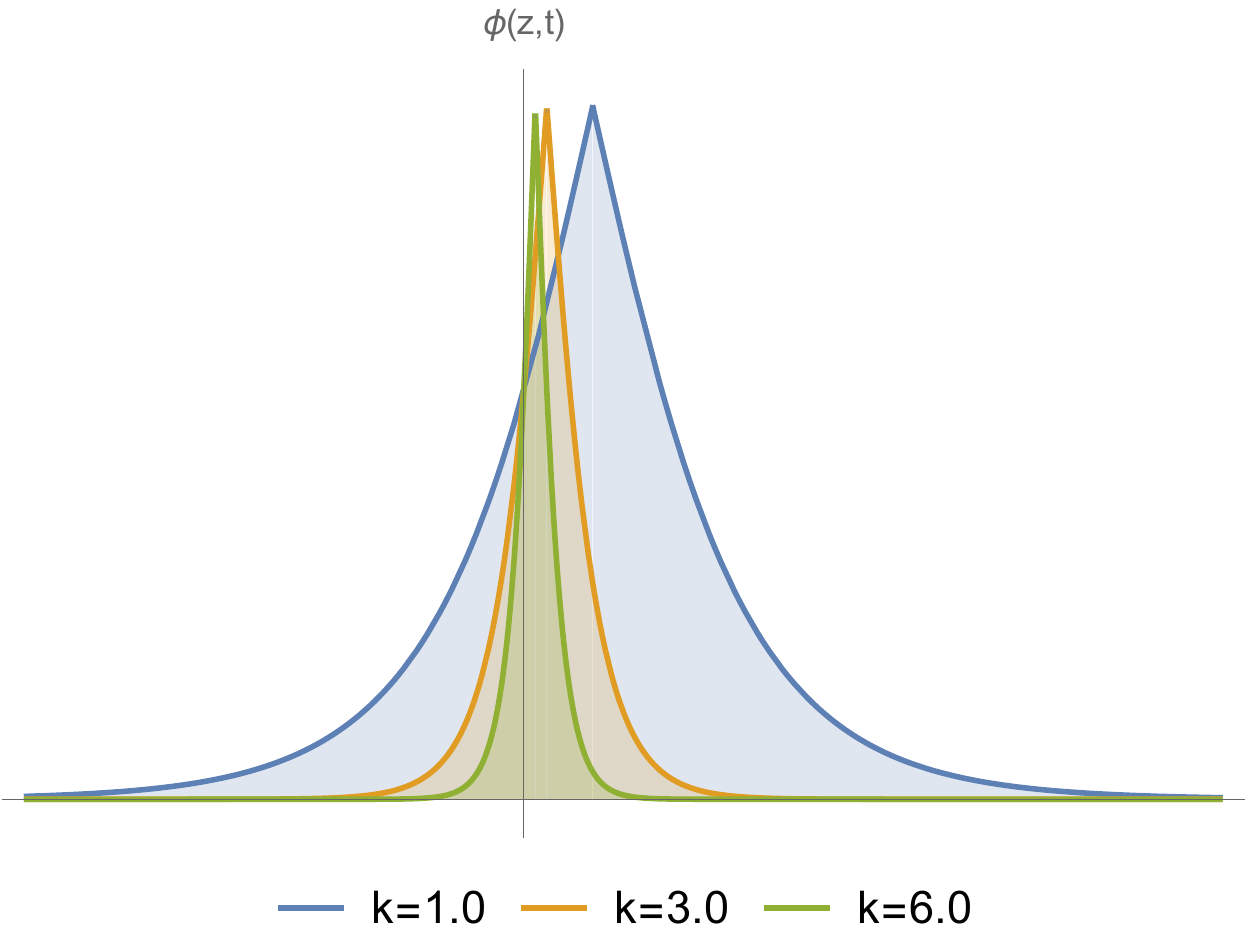}}
    &\parbox{3in}{\includegraphics[scale=0.6]{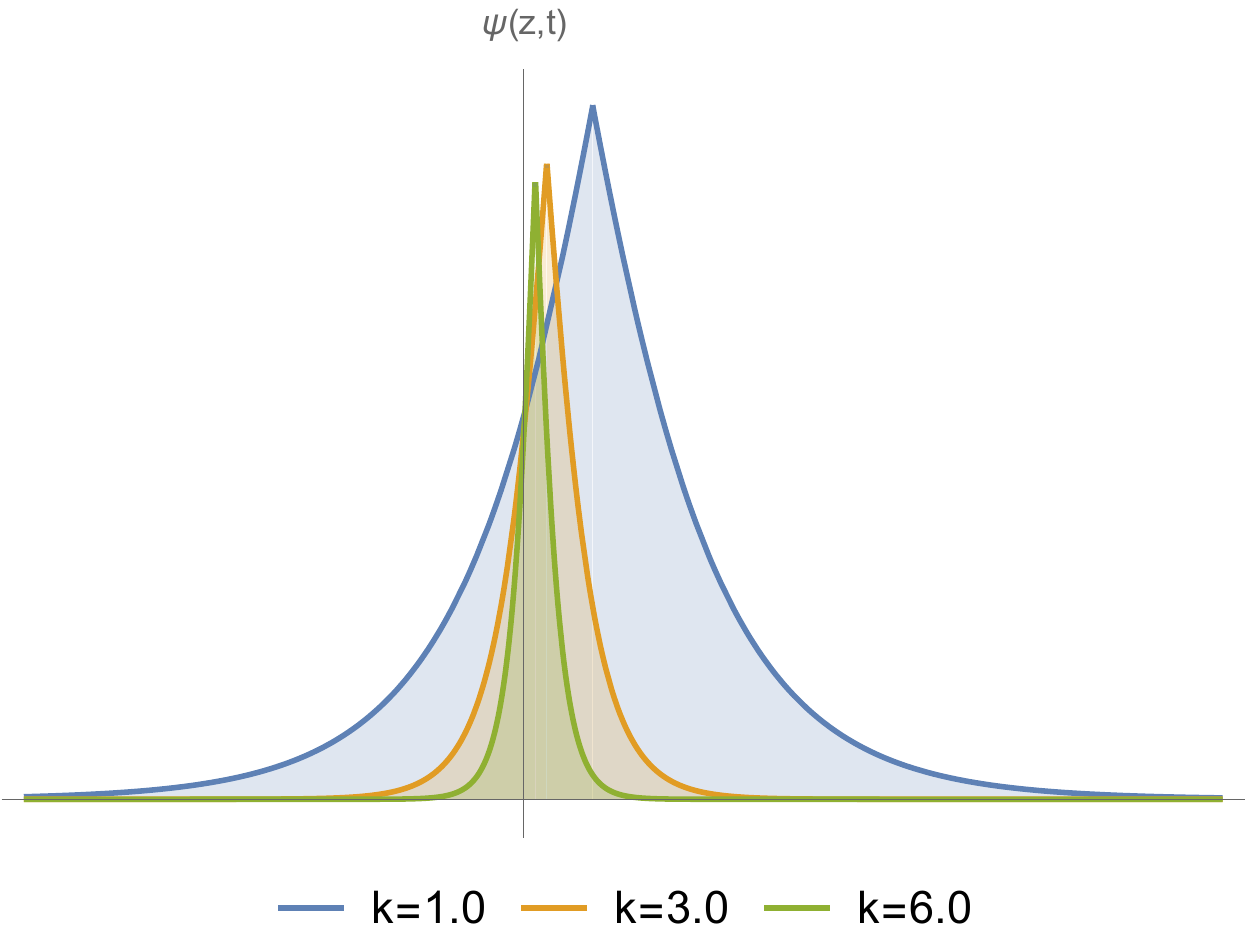}}\\
    \parbox{2in}{\includegraphics[scale=0.4]{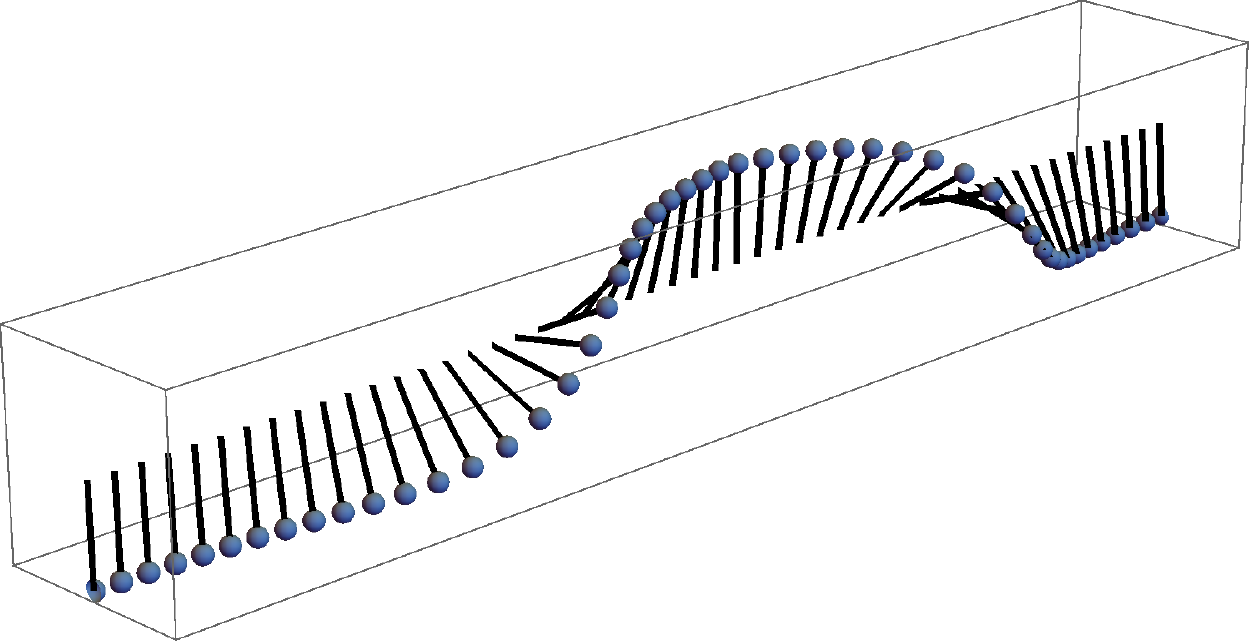}}
    &\parbox{2.1in}{\includegraphics[scale=0.45]{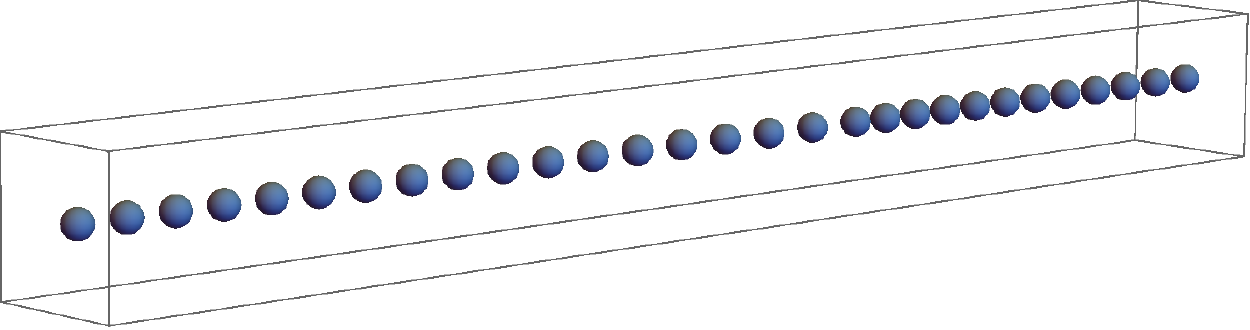}}\\
    \parbox{2in}{\includegraphics[scale=0.4]{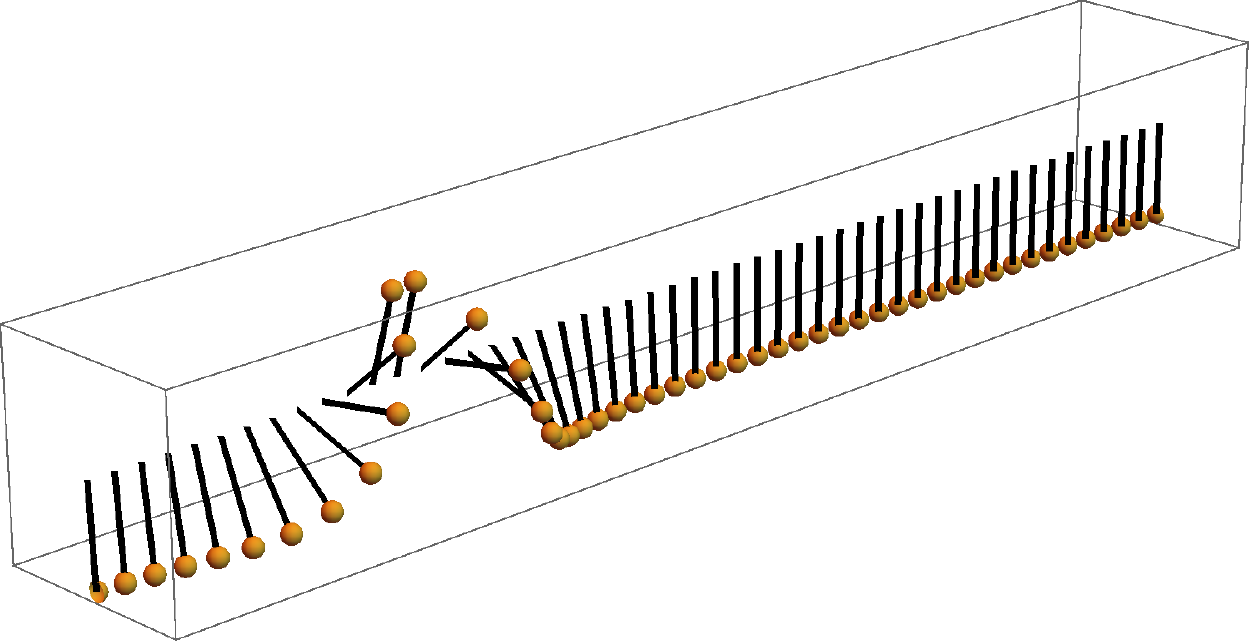}}
    &\parbox{2in}{\includegraphics[scale=0.45]{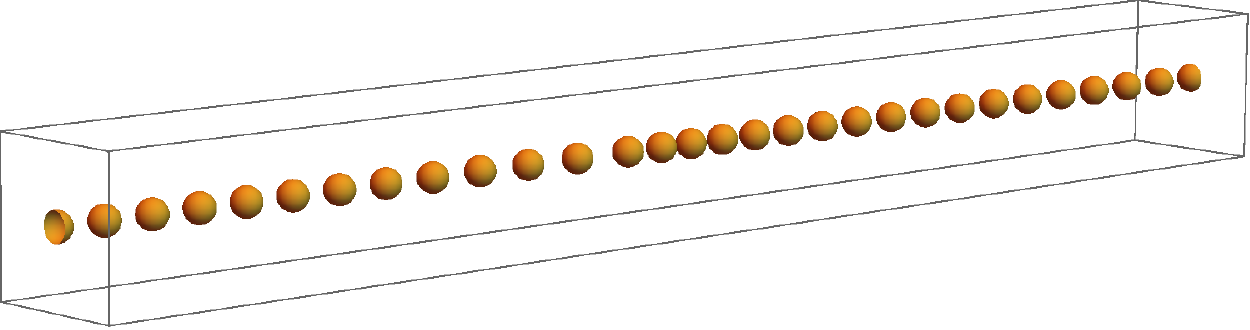}}\\
    \parbox{2in}{\includegraphics[scale=0.4]{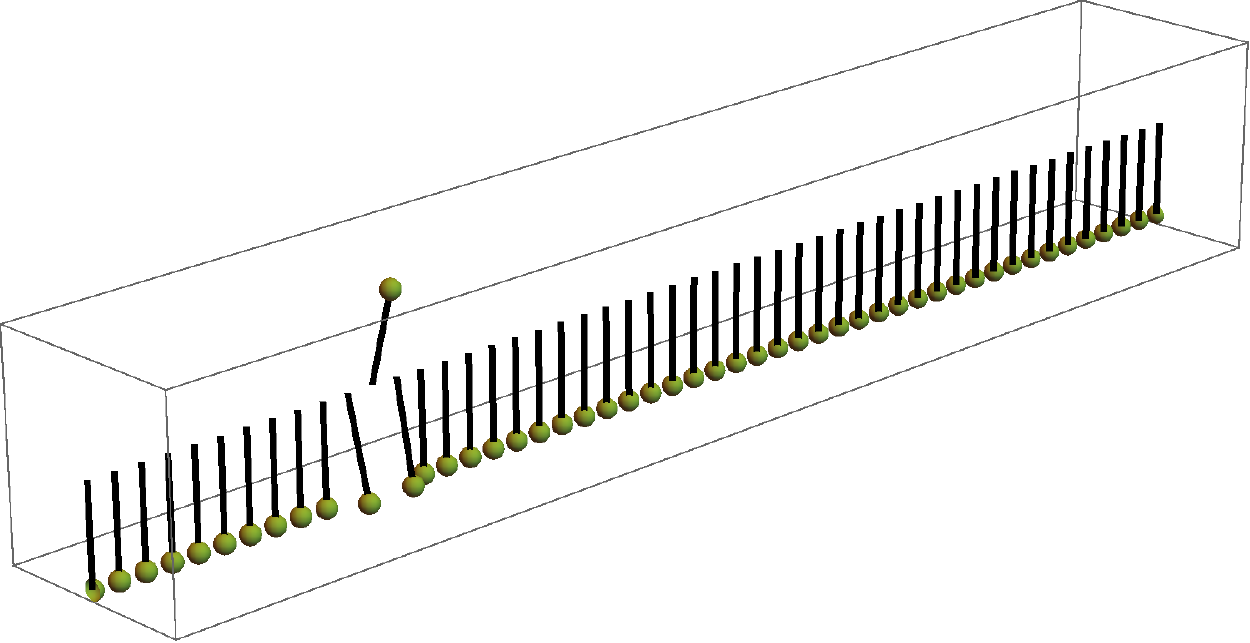}}
    &\parbox{2in}{\includegraphics[scale=0.45]{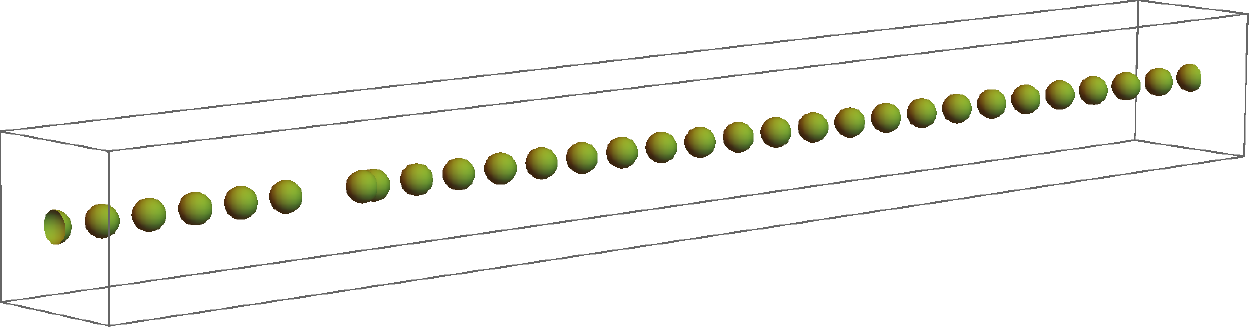}}
  \end{tabular}
  \caption{For small values of $k$ (the blue shaded wave, pendulums and beads), we observe the width of rotational/displacement deformation is broad, while we observe narrow rotational/displacement deformations for large values of $k$ (the green shaded wave, pendulums and beads). }
  \label{f003}
\end{figure}

In Fig.~\ref{f003}, the soliton solutions for $\phi(z,t)$ and $\psi(z,t)$ are given at $t=0$ with corresponding values of $k$. As the rotational wave $\phi(z,t)$ propagates with a speed $v$ along the $z$-axis, the points of micro-continuum (displayed as pendulums along the $z$-axis) experience microrotational deformations perpendicular to the axis. In the same way the longitudinal solution $\psi(z,t)$ gives rise to the compressional deformation wave propagating with the same speed $v$, on the points of macro-continuum (shown as beads) along the axis. As we vary the values of $k$, the widths of the soliton solutions are changed and this affects the overall deformational behaviours both in rotation and displacement.

\section{Properties of solutions}

We notice that there might be possible singularity issues in the amplitude of $\psi(z,t)$ in (\ref{3.37}) as $v^2$ approaches  $v^2_{\text{elas}}$. In order to resolve this problem, we would like to look closely at $k$ as a function of $v$ taking account of all nine parameters, $\{\kappa_1,\kappa_3,\chi_1,\chi_3,\rho,\rho_{\text{rot}},\mu_c,\lambda,\mu\}$.
We consider only the positive roots of $k^2$ to understand the possible range of $k$ for a given $v$. After putting all relevant parameters in (\ref{3.33}), we obtain
\begin{equation}\label{3.41}
  \begin{split}
    k&=3\left(\frac{\lambda^2+(\lambda+2\mu-v^2\rho)\mu_c}{3(\lambda+2\mu-v^2\rho)(\kappa_1+6\kappa_3)-9v^2\rho_{\text{rot}}(\lambda+2\mu-v^2\rho)-(3\chi_1-\chi_3)^2}\right)^{1/2}\\
    &=\frac{3}{\sqrt{\rho\rho_{\text{rot}}}}\left(\frac{\lambda^2+\mu_c\rho(v^2_{\text{elas}}-v^2)}{v^4-(v^2_{\text{elas}}+v^2_{\text{rot}})v^2+(v^2_{\text{elas}}v^2_{\text{rot}}-M_{12}M_{21})}\right)^{1/2}.
  \end{split}
\end{equation}
Now, to determine whether $k$ possesses any singularity, we compute the discriminant of the quartic of $v$ in the denominator regarding it as a quadratic equation for $v^2$.
\begin{equation}\label{3.42}
  (v^2_{\text{elas}}+v^2_{\text{rot}})^2-4(v^2_{\text{elas}}v^2_{\text{rot}}-M_{12}M_{21})=(v^2_{\text{elas}}-v^2_{\text{rot}})^2+\frac{16\rho_{\text{rot}}}{\rho}v^4_\chi
\end{equation}
where we put $v^2_\chi\equiv M_{12}$. This is strictly non-negative, so that we can have four roots of $v$ in the denominator of (\ref{3.41}), which will cause the singularity of $k$. We denote the four distinct roots as $v_i$, $i=1,2,3,4$ and assume that $v_1<v_2<0<v_3<v_4$. In particular, we write explicitly
\begin{equation}\label{3.43}
  v^2=\frac{1}{2}\left((v^2_{\text{elas}}+v^2_{\text{rot}})\pm\sqrt{(v^2_{\text{elas}}-v^2_{\text{rot}})^2+\frac{16\rho_{\text{rot}}}{\rho}v^4_\chi}\right)\;.
\end{equation}
The square root of this gives the four roots of $v_i$ where two positive roots $v_3$ and $v_4$ are related to two negative roots $v_1$ and $v_2$ by $v_3=-v_2$ and $v_4=-v_1$.

It can be recognised immediately that the values of $v_\text{elas}$ and $v_{\text{rot}}$ are restricted by
\begin{equation*}
  v_1\le-v_{\text{elas}},-v_{\text{rot}}\le v_2\qquad\text{and}\qquad v_3\le v_{\text{elas}},v_{\text{rot}}\le v_4.
\end{equation*}
Also, we will have $k=0$ if $v$ becomes 
\begin{equation}\label{3.44}
  v_0^2\equiv(\lambda^2/\rho\mu_c)+v^2_{\text{elas}}.
\end{equation}

Now, we plot the profiles of $v$ as a function of $k$, this is given implicitly by (\ref{3.41}), and we consider only the positive values of $v$ for the simplicity. At this time, we only have two asymptotic lines of $v_3$ and $v_4$ (again we assume $v_3<v_4$). And we assume that $v_{\text{elas}}>v_{\text{rot}}$.

\begin{figure}[!htb]
  \begin{tabular}{cc}
    \parbox{0.5\textwidth}{\includegraphics[width=0.45\textwidth]{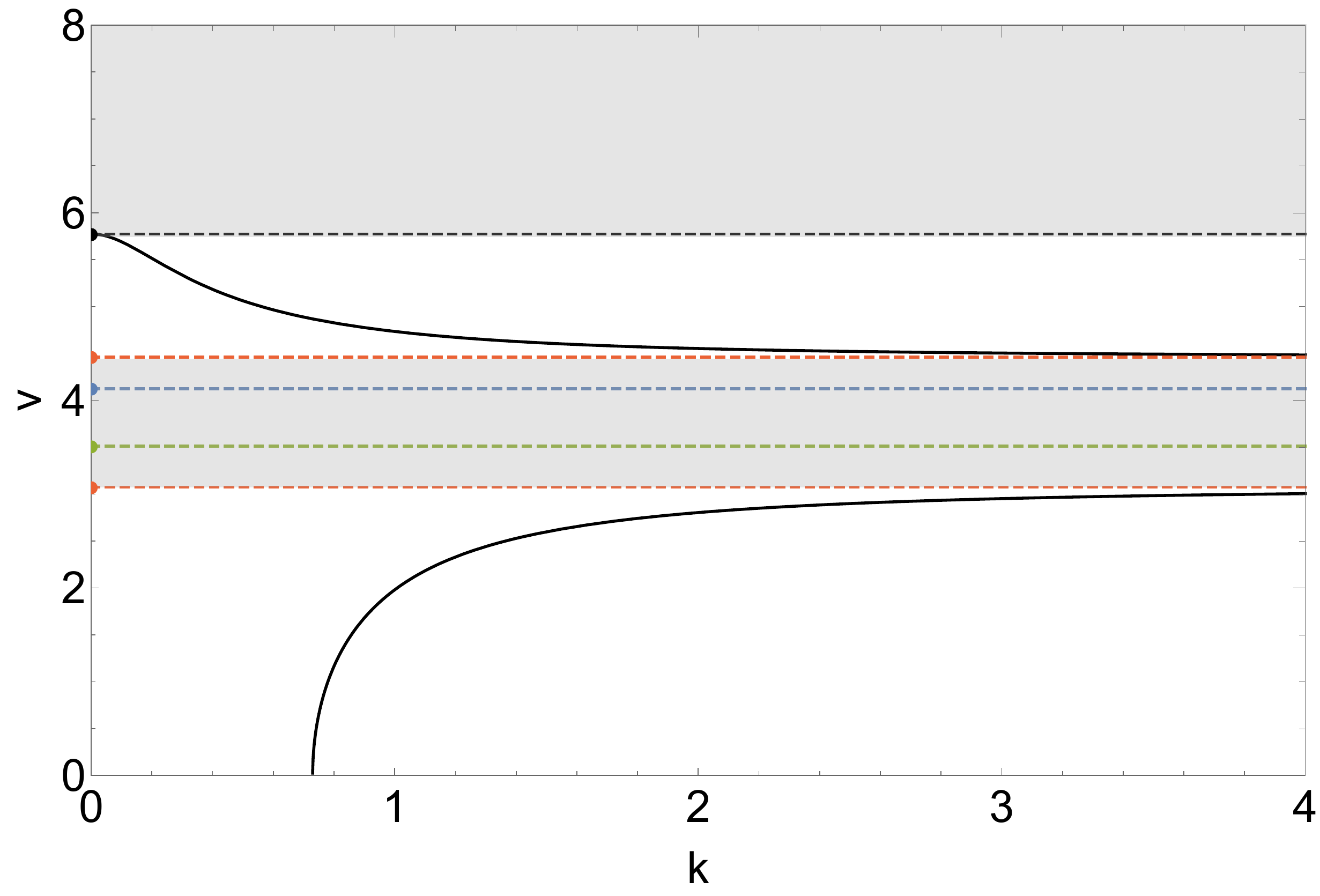}}
    &\parbox{0.5\textwidth}{\includegraphics[width=0.45\textwidth]{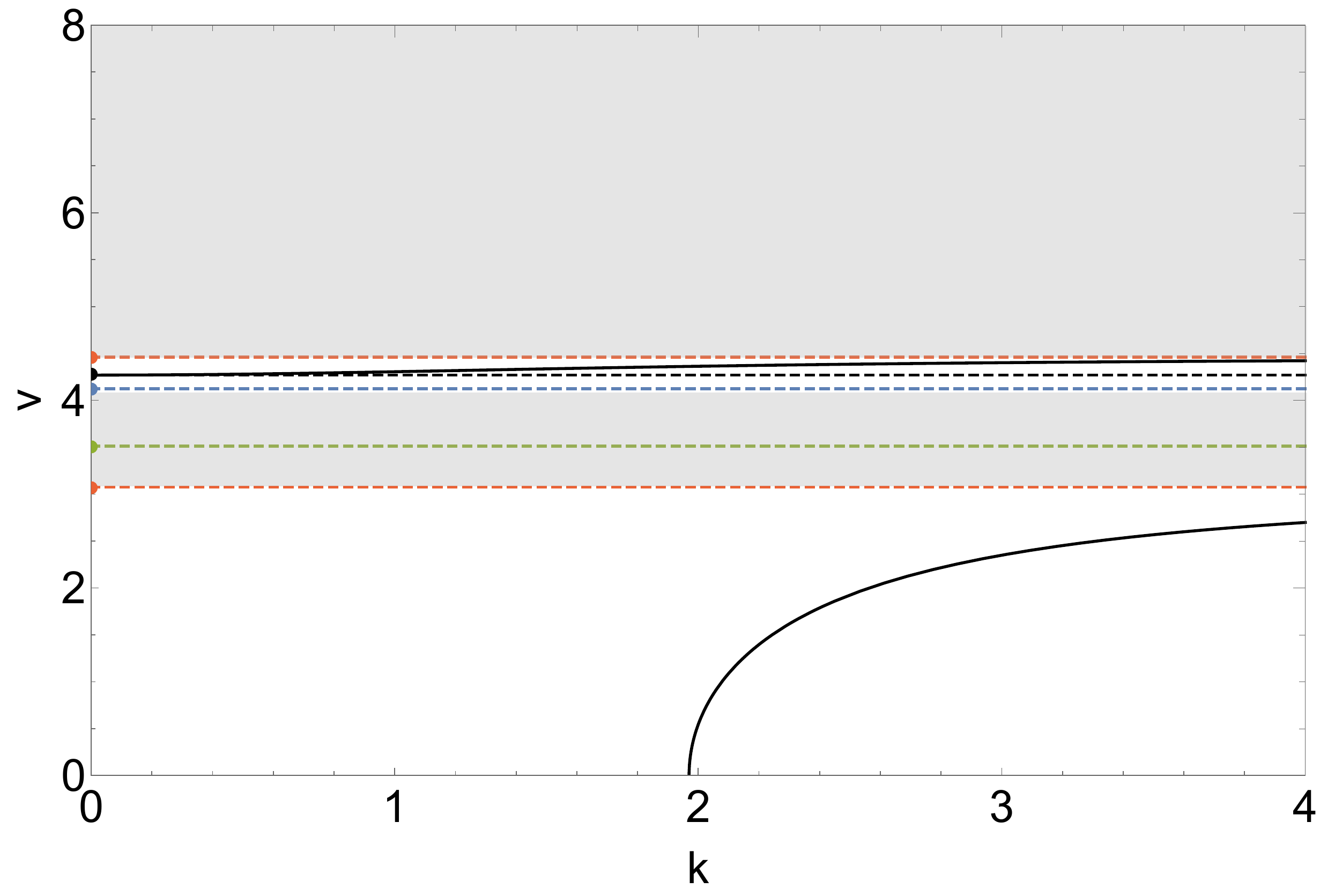}}\\[1ex]
    type $(a)$ & type $(b)$
  \end{tabular}
  \caption{The dashed lines indicate the position of $v_0$, the dot-dashed lines are for $v_{\text{elas}}$ and $v_{\text{rot}}$. The positions of asymptotic lines  $v_3$, $v_4$ are shown in dotted lines. We put the values of parameters $(\kappa_1,\kappa_3,\chi_1,\chi_3,\rho,\rho_{\text{rot}},\mu_c,\lambda,\mu)=(0.7,0.5,0.5,0.1,0.1,0.1,0.3,1.0,0.5)$ for type $(a)$. For type $(b)$, we alter one value of parameters $\mu_c=1.2$. In this way, we obtain two distinct types of behaviours of $v$ and $k$. This again determines two characteristic overall behaviours of the soliton solution of Fig.~\ref{f003}.}
  \label{f004}
\end{figure}

Two characteristic types of parameter ranges for $v$ with various values for a set of parameters with relevant asymptotic lines and the locations of $v_0$, $v_{\text{elas}}$ and $v_{\text{rot}}$ are given in Fig.~\ref{f004}. The dominating set of parameters in determining the characteristics is the set of constants $\{\lambda,\mu,\mu_c\}$ of the energy functional $V_{\text{elastic}}$. Notably, we observe that we only alter the value of the parameter $\mu_c$ to obtain the type $(b)$ solution from the type $(a)$ solution while keeping all remaining parameters unchanged. The values of $v_{\text{elas}}$ and $v_{\text{rot}}$ are located inside (or on the boundary of) the shaded region surrounded by asymptotic lines, which can be shown directly from (\ref{3.43}). The threshold in transition from the type $(a)$ to $(b)$ is evidently the relative positions between $v_0$ and $v_4$. If $v_0>v_4$ we will have the type $(a)$ and if $v_0<v_4$ then the type $(b)$.

\begin{figure}[!htb]
  \includegraphics[width=0.45\textwidth]{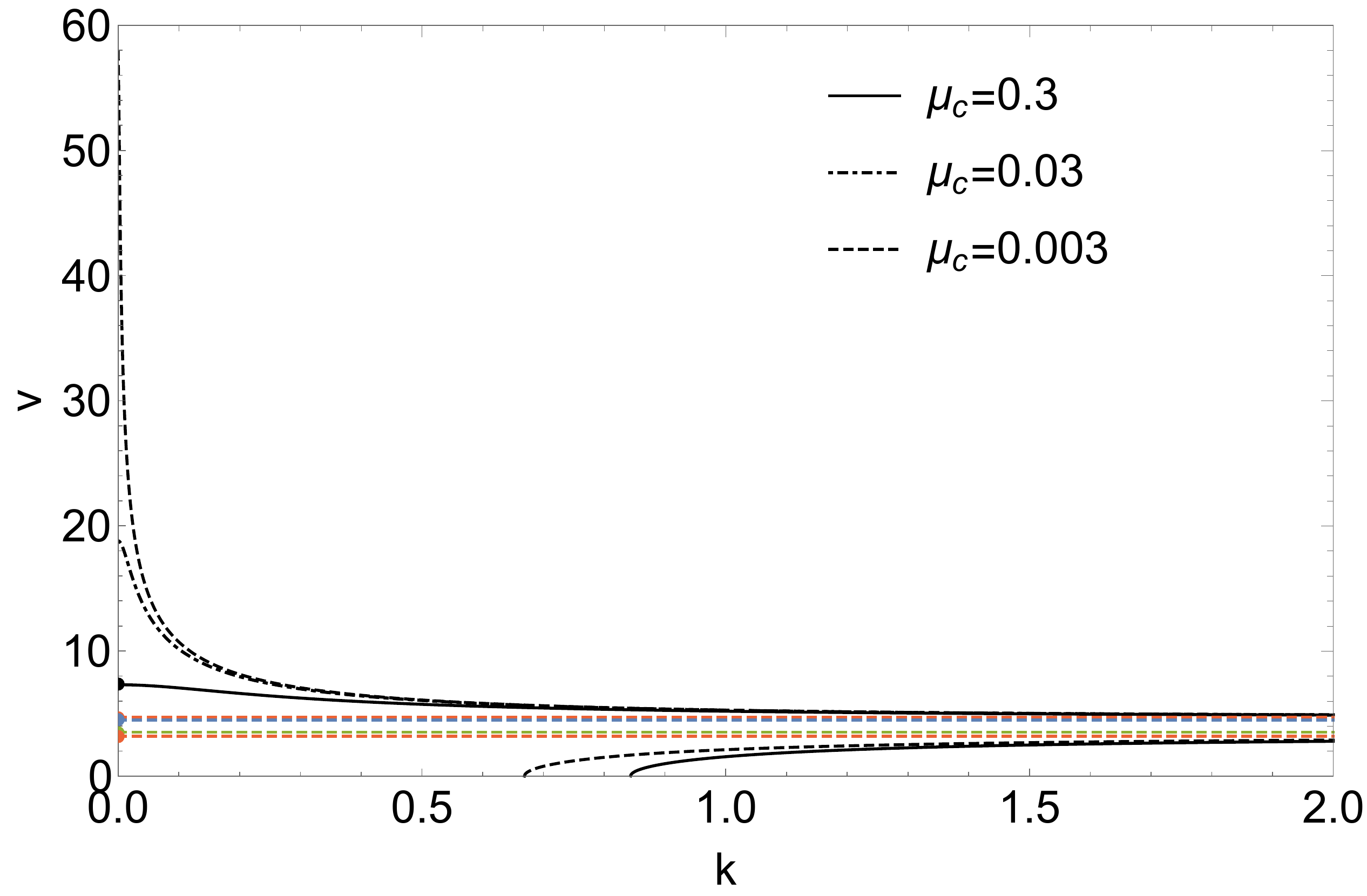} \hfill
  \includegraphics[width=0.45\textwidth]{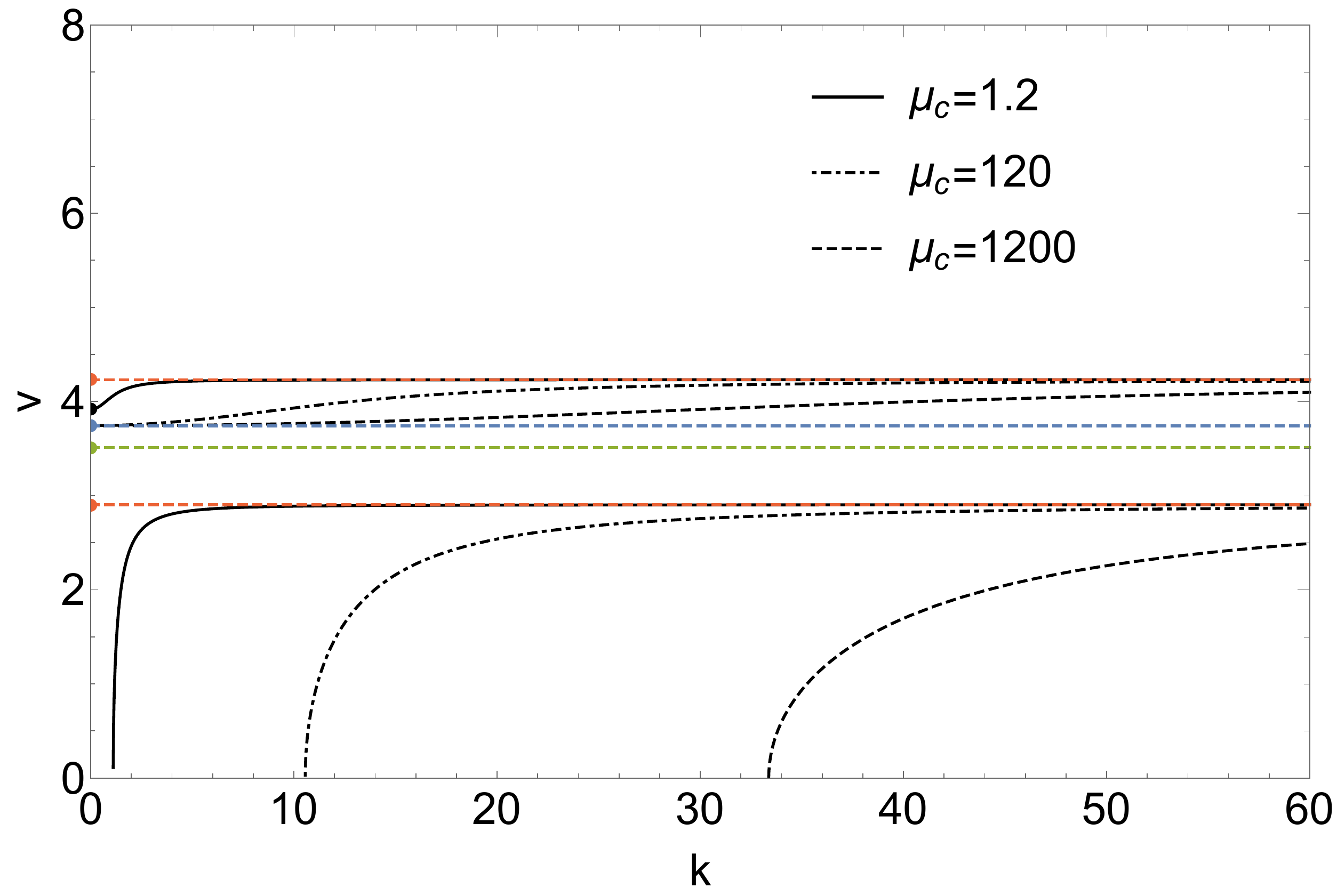}
  \caption{We indicate the modified profiles of the type (a) and (b) solutions as dot-dashed and dotted lines in the two limits of $\mu_c\to 0$ and $\mu_c\to\infty$. As $\mu_c\to 0$, the upper boundary $v_0$, in the type (a) of Fig.~\ref{f004}, is pushed up to the infinity (left).  In the limit $\mu_c\to\infty$ the lower profile of type (b) will shift to infinity along the $k$ axis (right).}
  \label{f006}
\end{figure}

In both types $(a)$ and $(b)$ solutions, there exist regions (the shaded regions) in which $v$ cannot be defined for a given $k$, solutions with such parameter choices do not exist. In case of type $(a)$, the values of $v$ are defined in $v\in[0,v_3)$ and $v\in(v_4,v_0]$. The upper limit of $v$ is bounded by $v_0$ and we can see that $v_0\to\infty$ as $\mu_c\to 0$ which is evident from (\ref{3.44}), see Fig.~\ref{f006}.

On the other hand, for the type $(b)$, the position of $v_0$ is $v_3<v_{\text{elas}}<v_0<v_4$. Now, the line of $v_0$ acts the role of the boundary line along with $v_3$ in $(b)$. So $v$ takes the values in the region $v\in[0,v_3)$ and $v\in[v_0,v_4)$. We must notice that for type $(b)$ solutions, the value of $v_0$ cannot be exactly $v_{\text{elas}}$ due to the restriction (\ref{3.44}), as long as we have nonzero $\lambda$. We observe that $v_0$ approaches $v_{\text{elas}}$ as $\mu_c\to\infty$, but the lower profile of $v$ in $(b)$ will be shifted to the right indefinitely, i.e. $k\to\infty$, see Fig.~\ref{f006}. In the limit $\mu_c\to\infty$, it is clear that we will have a profile of type (b). Also we can see from (\ref{3.33}) that $m^2\to m^2_0$, hence $k^2\to k^2_0$. This suggests that $b$ becomes negligible and we will be left with the soliton solution $\phi\to\phi_0$ of the form (\ref{3.27}).
    
\begin{figure}[!htb]
  \begin{tabular}{cc}
    \parbox{0.45\textwidth}{\includegraphics[width=0.45\textwidth]{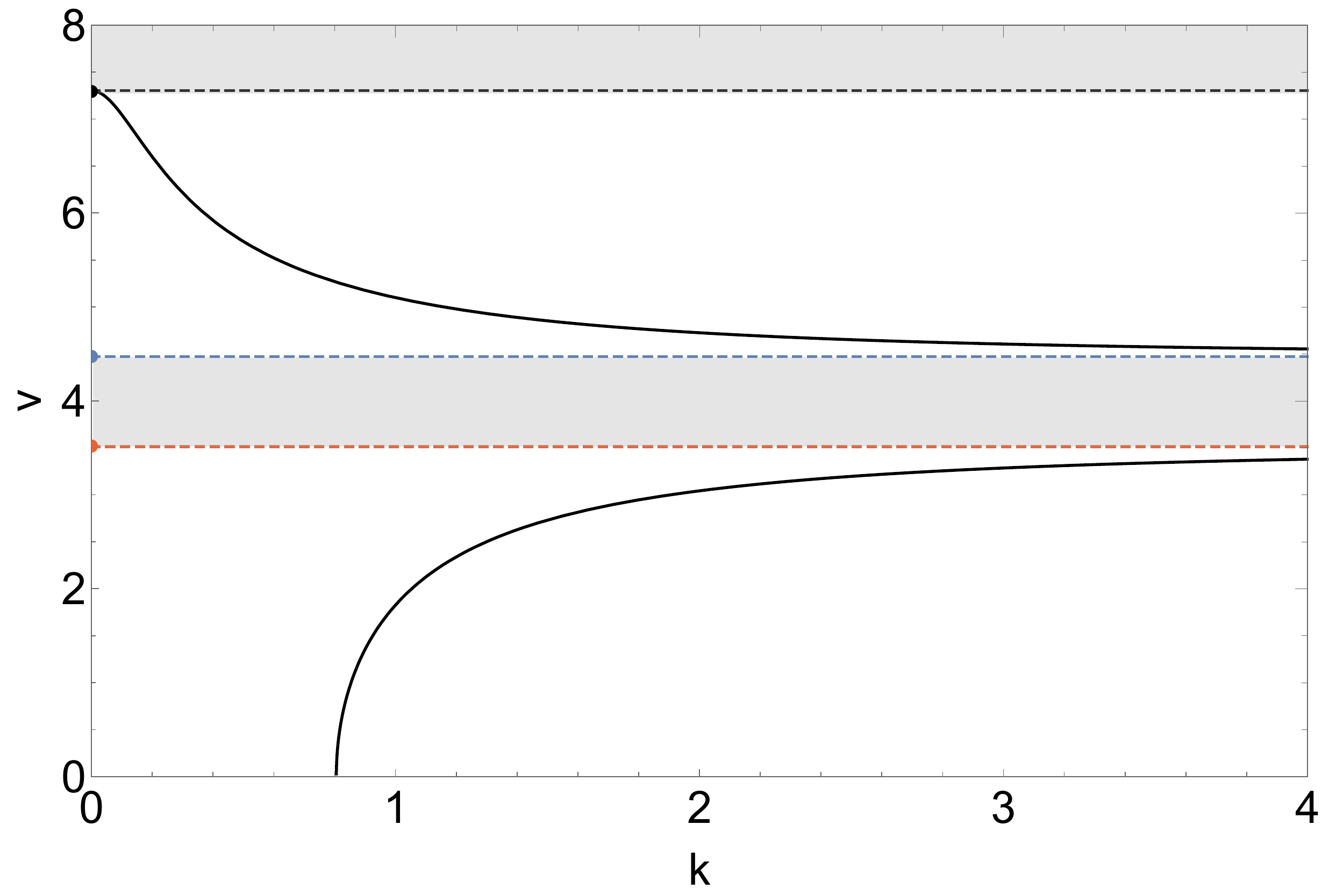}}
    &\parbox{0.45\textwidth}{\includegraphics[width=0.45\textwidth]{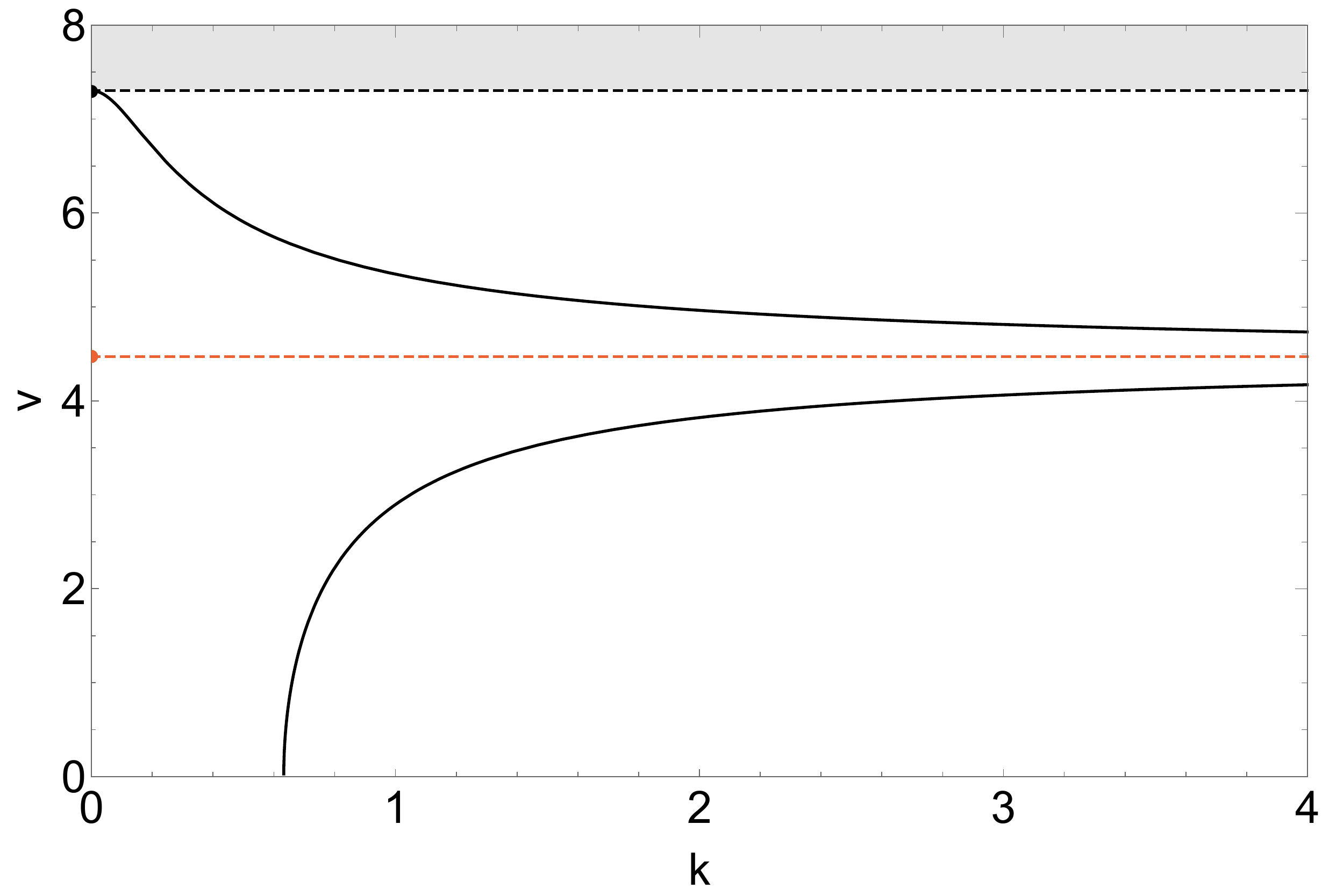}}\\[1ex]
    type $(c)$ & type $(d)$
  \end{tabular}
  \caption{For $(c)$, we put $(\kappa_1,\kappa_3,\chi_1,\chi_3,\rho,\rho_{\text{rot}},\mu_c,\lambda,\mu)=(0.7,0.5,0.5,1.5,0.1,0.1,0.3,1.0,0.5)$ so that $3\chi_1-\chi_3=0$ and we obtain $v_4=v_{\text{elas}}=4.47214$ and $v_3=v_{\text{rot}}=3.51188$. For $(d)$, we only altered value of parameter $\kappa_1=3.0$ so that $v_{\text{elas}}=v_{\text{rot}}=v_3=v_4=4.47214$.}
  \label{f005}
\end{figure}

Next, we consider the limit
\begin{equation}\label{3.45}
  \frac{\rho_{\text{rot}}}{\rho}\frac{v^4_\chi}{(v^2_{\text{elas}}-v^2_{\text{rot}})^2}\ll 1.
\end{equation}
In this limit, we can approximate the expressions of $v_3$ and $v_4$ given by (\ref{3.43}) as follows
\begin{equation}\label{3.46}
  v_4\approx v_{\text{elas}}\left(1+\frac{2\rho_{\text{rot}}v^4_\chi}{\rho(v^2_{\text{elas}}-v^2_{\text{rot}})v^2_{\text{elas}}}\right), \quad\quad
  v_3\approx v_{\text{rot}}\left(1-\frac{2\rho_{\text{rot}}v^4_\chi}{\rho(v^2_{\text{elas}}-v^2_{\text{rot}})v^2_{\text{rot}}}\right).
\end{equation}
Hence we can see that $v_{\text{elas}}$ approaches to $v_4$ and $v_{\text{rot}}$ approaches to $v_3$ for the type $(a)$ parameter choice. In case of type $(c)$ of Fig.~\ref{f005}, we set $v_\chi=0$ (i.e., $3\chi_1-\chi_3=0$) to illustrate that $v_{\text{elas}}=v_4$ and $v_{\text{rot}}=v_3$ and that the lines of $v_{\text{elas}}$ and $v_{\text{rot}}$ play the role of asymptotic lines. In this case, the matrix $\B{M}$ of (\ref{3.13-2}) becomes diagonal and the system looks similar to (\ref{3.13-3}). Of course, if we had assumed that $v_{\text{elas}}<v_{\text{rot}}$, then we would have $v_{\text{rot}}=v_4$ and $v_{\text{elas}}=v_3$. We may obtain the similar observation in type $(b)$ diagram by adjusting $\mu_c$, but $v_{\text{elas}},v_{\text{rot}}\to v_3$. Furthermore, in the same limit of $v_\chi=0$, if we set an additional condition that $v_{\text{elas}}=v_{\text{rot}}$, then we will have one asymptotic line $v_{\text{elas}}$ as shown in the diagram, type $(d)$ of Fig.~\ref{f005} and the matrix $\B{M}$ will simply become the identity matrix (up to the rescaling).

Now, the amplitude of $\psi$ in (\ref{3.37}) is determined by two coefficients (the matrix element $M_{21}$ can be written in terms of $v^2_\chi\equiv M_{12}$),
\begin{equation}
  \label{3.46a}
  \frac{16\rho_{\text{rot}}v^2_\chi}{\rho(v^2-v^2_{\text{elas}})}\qquad\text{and}\qquad\frac{4\lambda}{\rho k(v^2-v^2_{\text{elas}})}\;.
\end{equation}
The analytic investigation on the profiles of $v$ as a function of $k$ provides us the clue that the amplitude of $\psi$ cannot be arbitrarily large. As $k\to\infty$, we have $v^2\to v^2_{\text{elas}}$ but the statement that the value of $v^2_{\text{elas}}$ approaches $v^2_4$ is equivalent to say that $v^2_\chi\to 0$, as we can see directly from (\ref{3.46a}). Hence the first coefficient in~(\ref{3.46a}) is assumed to remain finite in this limit. Similarly, the second coefficient cannot be arbitrarily large. For given $k$ and $(v^2-v^2_{\text{elas}})$ will compensate each other as $k\to\infty$. This is shown in the type $(c)$, or more extreme case, the type $(d)$ in Fig.~\ref{f005}. 

\section{Conclusion}

We extended the previous study of the deformations considered in \cite{CB2016-2} to include the fully nonlinear model with arbitrarily large rotations and displacements. This discussion gave us further insights into the nature of the nonlinear geometry of Cosserat micropolar elasticity. The solution $\phi$ differs from $\phi_0$ via the different form of $k$ in (\ref{3.33}). On the other hand, the displacements $\psi$ and $\psi_0$ differ by additional nonlinear terms.

The soliton solutions for both rotations and displacements were obtained from the equations of motion and these allowed us to understand the geometric interpretation of the deformation waves. The physically dominant parameters of the complete model are the Lam\'{e} parameters $\{\lambda,\mu\}$ and the Cosserat couple modulus $\mu_c$. This becomes evident by looking at the $k$ dependency, or equivalently $m$ dependency, of the soliton solutions on these parameters.

The various values for $k$ in the soliton solutions for $\phi$ and $\psi$ give different overall behaviours while other values of parameters are fixed. Regarding the microrotations, the effect becomes apparent for large values of $k$, which induce high-frequency of localised energy distribution on the narrow width affected cross section both for the rotational and displacement deformations, whereas small values of $k$ induce gradual and broad energy distribution for the deformations over the microcontinuum media. The role of $k$ can be understood using a simple model of beads and pendulums as shown in Fig.~\ref{f003}.

A consideration for the deformation waves of higher dimensions would be a natural extension of the procedure. Some other candidates for further applications would include an investigation of domain walls in topological defects (e.g. ferromagnets) in connection with micropolar deformation. Moreover vortices as topological solitons with a notion of spontaneous symmetry breaking as a phase transition by Cosserat elasticity would be also be an interesting subject of study.

\section*{Acknowledgement}

Yongjo Lee is supported by EPSRC Doctoral Training Programme (EP/N509577/1). We would like to thank Sebastian Bahamonde who contributed to computing the equations of motion. 

\appendix
\section{Variations of energy functional}
\label{sec:app}

We would like to vary each energy functional using some of identities listed in Notation and Appendix. First, for $V_{\text{elastic}}$, we can expand the expression using the definition of $\left\|X\right\|^2=\langle X,X\rangle=\tr(XX^T)$ and $\text{sym}M=1/2(M+M^T)$ as
\begin{multline}\label{a2.5}
  V_{\text{elastic}}(F,\overline{R})=
  \mu\left\|\text{sym}\overline{R}^TF-\id\right\|^{2}+
  \frac{\lambda}{2}\Bigl[\tr(\text{sym} (\overline{R}^{T}F)-\id\Bigr]^{2}\\
  =\left(3\mu+\frac{9}{2}\lambda\right)+
  \frac{1}{2}\mu\tr\left(\overline{R}^TF\overline{R}^TF\right)+\frac{1}{2}\mu\tr(FF^{T})
  -\left(2\mu+3\lambda\right)\tr(\overline{R}^TF)+
  \frac{\lambda}{2}\left[\tr(\overline{R}^{T}F)\right]^2.
\end{multline}
Variation of this is
\begin{equation}\label{a2.6}
  \begin{split}
    \delta V_{\rm elastic}(F,\overline{R})&=\Big[\mu(\overline{R}F^{T}\overline{R}+F)-(2\mu+3\lambda)\overline{R}+ \lambda \tr(\overline{R}^{T}F)\overline{R}\Big]:\delta F\\
    &\qquad+\Big[\mu F\overline{R}^{T}F-(2\mu+3\lambda)F+\lambda\tr(\overline{R}^{T}F)F\Big]:\delta \overline{R}.
  \end{split}
\end{equation}
If we want to study the dynamical problem, we must take the kinetic term into account in the elastic energy functional.
\begin{equation}\label{a2.7}
  V_{\text{elastic,kinetic}}=\frac{1}{2}\rho\|\dot{\varphi}\|^2
\end{equation}
where $\rho$ is the constant density and $\varphi$ is the deformation vector. If we vary this term we will obtain
\begin{equation}\label{a2.8}
  \delta V_{\text{elastic,kinetic}}=-\rho\ddot{\varphi}\;\delta\varphi.
\end{equation}
But, since $\nabla\varphi=\id+\nabla u$ implies $\delta\varphi=\delta u$ and $\ddot{\varphi}=\ddot{u}$, the variation of elastic kinetic term can be rewritten as $\delta V_{\text{elastic,kinetic}}=-\rho\ddot{u}\;\delta u$ and the variation of dynamical expression for the elastic energy functional becomes
\begin{equation}\label{a2.10}
  \begin{split}
    \delta V_{\rm elastic}(F,\overline{R})&=\Big[\mu(\overline{R}F^{T}\overline{R}+F)-(2\mu+3\lambda)\overline{R}+ \lambda \tr(\overline{R}^{T}F)\overline{R}\Big]:\delta F\\
    &\qquad+\Big[\mu F\overline{R}^{T}F-(2\mu+3\lambda)F+\lambda\tr(\overline{R}^{T}F)F\Big]:\delta \overline{R}+\rho\ddot{u}\;\delta u.
  \end{split}
\end{equation}

Similarly, for the curvature functional, we can expand it as
\begin{equation}\label{a2.11}
  \begin{split}
    V_{\text{curvature}}(\overline{R})&=\frac{(\kappa_{1}-\kappa_{2})}{2}\tr\Big[\overline{R}^{T}(\Curl\overline{R})\overline{R}^{T}(\Curl\overline{R})\Big]\\
    &\qquad+\frac{(\kappa_{1}+\kappa_{2})}{2}\tr\Big[(\Curl\overline{R})^{T}(\Curl\overline{R})\Big]-\Big(\frac{\kappa_{1}}{3}-\kappa_{3}\Big)\Big(\tr\Big[\overline{R}^{T}(\Curl\overline{R})\Big]\Big)^2.
  \end{split}
\end{equation}
This is a functional dependent only on $\overline{R}$, but the actual variation will involve rather complicated quantites such as $\delta\Curl\overline{R}$ multiplied by a tensor. To overcome this problem, we introduce the following identity. Let $A(\overline{R})$ and $B(\overline{R})$ be two matrix valued functions depending on the rotation $\overline{R}$. Then, by direct calculation, one can show that an identity for any rank-two tensors $A$ and $B$,
\begin{equation}\label{a2.12}
  \tr(A)B :\delta(\Curl\overline{R}) = -\Big[B\big(\grad \tr(A)\big)^{\star}\Big]:\delta\overline{R}+\tr(A)\Curl B :\delta\overline{R}
\end{equation}
where
\begin{equation}\label{a2.13}
  \big(\grad \tr(A)\big)^{\star}_{ik} = \epsilon_{ijk} \partial_j \tr(A).
\end{equation}
The identity (\ref{a2.12}) can be shown if one uses the convention $\Curl B=\epsilon_{jrs}B_{is,r}e_{i}\otimes e_{j}=-B_{,i}\times e_{i}$. In particular, if we put $A=\id$ then (\ref{a2.12}) reduces to
\begin{equation}\label{a2.14}
  B:\delta(\Curl\overline{R})=\Curl B:\delta\overline{R}.
\end{equation}
And this will play an important role in simplifying the calculation of variation of the energy functionals significantly. For example, the first variational term in (\ref{a2.11}) would be
\begin{equation}\label{a2.15}
  \begin{split}
    \delta\left(\tr\Big[\overline{R}^{T}(\Curl\overline{R})\overline{R}^{T}(\Curl\overline{R})\Big]\right)&=2\Big[\overline{R}(\Curl\overline{R})^{T}\overline{R}\Big]:\delta( \Curl\overline{R})+2(\Curl\overline{R})\overline{R}^{T}(\Curl\overline{R}):\delta\overline{R}\\
    &=2\left(\Curl\Big[\overline{R}(\Curl\overline{R})^{T}\overline{R}\Big]+(\Curl\overline{R})\overline{R}^{T}(\Curl\overline{R})\right):\delta\overline{R}.
  \end{split}
\end{equation}
In this way, we find the variation of curvature term
\begin{multline}
    \delta V_{\text{curvature}}(\overline{R})=\Big[(\kappa_{1}-\kappa_{2})\Big((\Curl\overline{R})\overline{R}^{T}(\Curl(\overline{R}))+\Curl\Big[\overline{R}(\Curl\overline{R})^{T}\overline{R}\Big]\Big)+(\kappa_{1}+\kappa_{2})\Curl\Big[\Curl\overline{R}\Big]\\
      -\left(\frac{\kappa_{1}}{3}-\kappa_{3}\right)\Big(4\tr(\overline{R}^T\Curl\overline{R})\Curl(\overline{R}) -2\overline{R}\Big(\grad\Big(\tr[\overline{R}^{T}\Curl\overline{R}]\Big)\Big)^{\star}\Big]:\delta\overline{R}.
    \label{a2.16}
\end{multline}
Again, for the dynamical case, we need to include the kinetic term defined as
\begin{equation}\label{a2.17}
  V_{\text{curvature,kinetic}}=\rho_{\text{rot}}\|\dot{\overline{R}}\|^2=\rho_{\text{rot}}\tr(\dot{\overline{R}}\ \dot{\overline{R}^{T}})
\end{equation}
with variational form given by $\delta V_{\text{curvature,kinetic}}=-2\rho_{\text{rot}}\ddot{\overline{R}}:\delta \overline{R}$. Therefore, the variation of dynamical expression for the curvature energy functional can be written as
\begin{multline}
  \delta V_{\text{curvature}}(\overline{R}) = 
  \Big[(\kappa_{1}-\kappa_{2})\Big((\Curl\overline{R})\overline{R}^{T}(\Curl(\overline{R}))+
    \Curl\Big[\overline{R}(\Curl\overline{R})^{T}\overline{R}\Big]\Big)+
    (\kappa_{1}+\kappa_{2})\Curl\Big[\Curl\overline{R}\Big] \\
    -\left(\frac{\kappa_{1}}{3}-\kappa_{3}\right)
    \Big(4\tr(\overline{R}^T\Curl\overline{R})\Curl(\overline{R}) -
    2\overline{R}\Big(\grad\Big(\tr[\overline{R}^{T}\Curl\overline{R}]\Big)\Big)^{\star}+
    2\rho_{\text{rot}}\ddot{\overline{R}}\Big]:\delta\overline{R}.
  \label{a2.18}
\end{multline}

For the interaction energy functional, we expand terms $\text{dev}\;\text{sym}(\overline{R}^{T}\Curl\overline{R})$ and $\text{dev}\;\text{sym}(\overline{R}^{T}F-\id)$ to write
\begin{align}
  V_{\text{interaction}}&=\left(\chi_{1}-\frac{\chi_{3}}{3}\right)\tr(\overline{R}^{T}\Curl\overline{R})\tr(\overline{R}^{T}F)
  \nonumber \\
    &\qquad+\frac{\chi_{3}}{2}\left(\tr\Big[(\Curl\overline{R})^{T}F\Big]+\tr\Big[\overline{R}^{T}(\Curl\overline{R})\overline{R}^{T}F\Big]\right).
    \label{a2.19}
\end{align}
The variation of this involves the quantity $\delta\Curl\overline{R}$ as in the case of $V_{\text{curvature}}$, so we use the identity (\ref{a2.14}) to obtain
\begin{multline}
  \delta V_{\text{interaction}}(F,\overline{R}) =\left\{\Big(\chi_{1}-\frac{\chi_{3}}{3}\Big)\Big(2\tr(\overline{R}^{T}F)\Curl\overline{R}+\tr(\overline{R}^{T}\Curl\overline{R})F-\overline{R}\Big[\grad\Big(\tr[\overline{R}^{T}F]\Big)\Big]^{\star}\Big)\right. \\
  +\left.\frac{\chi_{3}}{2}\Big(\Curl F+(\Curl\overline{R})\overline{R}^{T}F+F\overline{R}^{T}(\Curl\overline{R})+\Curl(\overline{R}F^{T}\overline{R})\Big)\right\}:\delta \overline{R}\\
  +\left\{\chi_{1}\tr(\overline{R}^{T}\Curl\overline{R})\overline{R}+\frac{\chi_{3}}{2}\Big(\Curl\overline{R}+\overline{R}(\Curl\overline{R})^{T}\overline{R}\Big)-\frac{\chi_{3}}{3}\tr(\overline{R}^{T}\Curl\overline{R})\overline{R}\right\}:\delta F.
  \label{a2.20}
\end{multline}
Lastly, we write the coupling energy functional as
\begin{equation}\label{a2.21}
  V_{\text{coupling}}(F,\overline{R})=\mu_c\left\|\overline{R}^{T}\text{polar}(F)-\id\right\|^2=2\mu_{c}(3-\tr[\overline{R}^{T}\text{polar}(F)]).
\end{equation}
We note that this depends on $\overline{R}$ and $R=\text{polar}(F)$, hence depends on $\overline{R}$ and $F$. Therefore, the variation of coupling energy functional is of the form
\begin{equation}\label{a2.22}
  \delta V_{\text{coupling}}(F,\overline{R})=-2\mu_{c}R:\delta\overline{R}-2\mu_{c}\left[\frac{\partial}{\partial F}\Big(\tr[\overline{R}^{T}R]\Big)\right]:\delta F.
\end{equation}
The term in the brackets in the second term can be written as
\begin{equation}\label{a2.23}
  \frac{\partial}{\partial F}\Big(\tr[\overline{R}^{T}R]\Big)=\Big(\frac{dR}{dF_{ml}}\Big):\frac{\partial }{\partial R}\Big[\tr(\overline{R}^{T}R)\Big]=\Big(\frac{dR}{dF_{ml}}\Big):\overline{R}
  =\frac{1}{\det(Y)}\Big[RY(R^{T}\overline{R}-\overline{R}^{T}R)Y\Big]
\end{equation}
where $Y=\tr(U)\id-U$. In the first step, we used the chain rule and in the second and last steps we used the identities given in Appendix. Then the variation of coupling energy becomes
\begin{equation}\label{a2.24}
  \delta V_{\text{coupling}}(F,\overline{R})=-2\mu_{c}\overline{R}:\delta\overline{R}-\frac{2\mu_{c}}{\det(Y)}\left[RY(R^{T}\overline{R}-\overline{R}^{T}R)Y\right]:\delta F.
\end{equation}

We list some useful matrix identities below.
\begin{alignat}{2}
  \frac{\partial}{\partial X}\tr(F(X)) &= [f(X)]^{T} &\qquad\qquad
  \frac{\partial}{\partial X}\tr(X)&= I\\
  \frac{\partial}{\partial X}\tr(XA)&= A^{T}&
  \frac{\partial}{\partial X}\tr(AXB) &= A^{T}B^{T}\\
  \frac{d}{dX}(\tr(XX^{T})) &= 2X&
  \frac{d}{dX}(\tr(XA)) &= A^{T}.
\end{alignat}
Here $f$ stands for the scalar derivative of $F$. Moreover
\begin{align}
\frac{d}{dX}(\tr(AXBX))&=A^{T}X^{T}B^{T}+B^{T}X^{T}A^{T}\\[1ex]
\frac{dg(R(F))}{F_{ml}}&=\tr\left[\frac{dR}{dF_{ml}}
  \left(\frac{dg(R)}{dR}\right)^{T}\right]=\frac{dR}{dF_{ml}}:\left(\frac{dg(R)}{dR}\right).
\end{align}

\bibliographystyle{unsrt}
\bibliography{references}

\end{document}